\documentclass[pre,preprintnumbers,amsmath,amssymb]{revtex4-1}
\usepackage{graphicx,color}
\usepackage{bm}
\usepackage{epsfig}

\providecommand{\red}[1]{\textcolor{red}{#1}}

\begin{document}

\title{Extreme value statistics in a continuous time branching process: a pedagogical primer}
\author{Satya N. Majumdar and Alberto Rosso}
\affiliation{LPTMS, CNRS, Univ. Paris-Sud, Universit\'e Paris-Saclay, 91405 Orsay, France}

\begin{abstract}
We study a continuous time branching process where an individual splits into two daughters with rate $b$ and dies with rate 
$a$, starting from a single individual at $t=0$. We show that the model can be mapped exactly to a random walk problem where 
the population size $N(t)$ performs a random walk on a positive semi-infinite lattice. The hopping rate of
this random walker out of a site 
labelled $n$ is proportional to $n$, i.e., the walker gets more and more `active' or `agitated' as it moves further and 
further away from the origin--we call this an `agitated random walk' (ARW). We demonstrate that this random walk problem is 
particularly suitable to obtain exact explicit results on the extreme value statistics, namely, on the distribution of the 
maximal population size $M(t)= \max_{0\tau\le t}[N(\tau)]$ up to time $t$. This extreme value distribution displays markedly 
different behaviors in the three phases: (i) subcritical ($b<a$) (ii) critical ($b=a$) and (iii) supercritical $(b>a)$. In 
the subcritical phase, $Q(L,t)$ becomes independent of time $t$ for large $t$ and the stationary distribution $Q(L, 
\infty)$ decays exponentially with increasing $L$. At the critical point $b=a$, the distribution again becomes independent 
of $t$ for large $t$, but the stationary distribution $Q(L,\infty)\sim 1/L^2$ has a power law tail for large $L$. For finite 
but large $t$, the distribution at the critical point exhibits a scaling form $Q(L,t)\sim f_c(L/{at})/L^2$ where the scaling 
function $f_c(z)$ has a nontrivial shape that we compute analytically. In the supercritical phase, the distribution $Q(L,t)$ 
has a `fluid' part that becomes independent of $t$ for large $t$ and a `condensate' part (a delta peak centered at 
$e^{(b-a)t}$) which gets disconnected from the `fluid' part and moves rapidly to $\infty$ as time increases. We also verify 
our analytical predictions via numerical simulations finding excellent agreement. Our results can be applied to characterize 
the time dependence of the size of the maximally infected population up to a fixed time during an epidemic spread.

\end{abstract}

\maketitle

\date{\today}

\section{Introduction}

Branching Brownian motion (BBM) is a well studied stochastic process with a multitude
of applications across disciplines: bacterial migration and population growth, spread of infectious diseases,
reaction-diffusion models, disordered systems, nuclear reactions, cosmic ray showers, to name a 
few~\cite{Fisher13,Harris63,Mckean75,Lindvall76,Pakes98,Bramson78,SF79,DFL86,Bailey87,TT92,DS88,GKCB98,BDS2008,
BD2009,BD2011,RMS14,RMS15,RMS15_2,FLRC2023,DMRZ13}. The BBM process can be described very simply using
the paradigmatic example of migration and population growth of bacterias.
Consider a bacterial colony that starts with a single bacteria at $t=0$.
The bacteria can diffuse and also reproduce through binary fission, i.e., it undergoes
a cell division and generates
two identical daughter cells. Each of the daughter cells can further 
diffuse and reproduce independently via binary fissions. This leads to an exponential growth
of the pupulation size in the colony. However, a bacterial cell can also die due
to the lack of nutrients. The competition between the growth driven by
cell divisions (branching) and death and in addition, the migration via diffusion leads to an interesting
stochastic process where both the population size $N(t)$ at time $t$, as well as the
spatial extent of the colony are random variables and fluctuate from one realization to another.
Characterizing the behaviors of these random variables provides important information
about the evolution of the bacterial colony. The dynamics takes place in continuous time
via the following simple rules. In a small time interval $\Delta t$,
a bacteria can do one of the three following moves

\begin{itemize}

\item With probability $b\, \Delta t$, a bacterial cell branches, i.e., divides
into two identical daughter cells.

\item With probability $a\, \Delta t$, the cell dies.

\item With the complementary probability $1-(b+a)\, \Delta t$, the cell diffuses with a diffusion
constant $D$.

\end{itemize}

Thus this simple model has three parameters $(a,b, D)$ and the process starts with a single bacteria at $t=0$.
Let $N(t)$ denote the population size at time $t$ which is a random variable. The first natural question
is: what is the probability distribution $P(n,t)= {\rm Prob.}[N(t)=n]$ of the population size at time $t$?
This quantity can be computed exactly (see e.g. Refs~\cite{RMS15} for a simple derivation that is
also reproduced here later in Section I). In particular, it is easy to show that the average population size
is given exactly at all times $t$ by the simple expression
\begin{equation}
\langle N(t)\rangle= e^{(b-a)\,t}\, .
\label{avg_Nt.1}
\end{equation}
Hence, for $b>a$, the average
population size grows exponentially with time: this is called
the `supercritical' phase ($b>a$). In contrast, for $b<a$, the average size vanishes exponentially
for large $t$ leading to an extinction of the bacterial colony: this
is the `subcritical' phase $b<a$. Exactly for $b=a$, the average size
$\langle N(t)\rangle=1$ does not change with time from its initial value $1$.
This is called the `critical' phase ($b=a$). The statistics of $N(t)$ is clearly
independent of the diffusion constant $D$ since diffusion plays no role on
the population size dynamics in this simple model. In the context of epidemic spread,
$b$ is the infection rate, $a$ denotes the recovery or death rate and $N(t)$ represents
the number of infected individuals at time $t$.

The spatial properties of the process have also been studied for a long time. For example, in one dimension, one 
can take a snapshot of the process at a fixed time $t$ and the ordered positions of the particles (whose total 
number $N(t)$ is random) $\{x_1(t)< x_2(t),\ldots, <x_{N(t)}(t)\}$ form a point process on a line. One can ask 
various interesting questions on the order statistics of this point process, e.g., the distribution of the 
position of the rightmost particle, the statistics of gaps between two consecutive particles etc. For example, for 
a=0, the cumulative distribution of the position of the rightmost particle is known to satisfy the 
Fisher-Kolomogorov-Petrovskii-Piskunov (Fisher-KPP) equation 
~\cite{Mckean75,Bramson78,GKCB98,BDS2008,BDS2008,BD2009,BD2011} with the typical value growing linearly with $t$. 
For $a=0$, the gap statistics near the rightmost (or leftmost) particle has also been 
studied~\cite{BD2009,BD2011}, though they are harder to compute analytically. At the critical point $b=a$, 
however, the statistics of the rightmost particle as well as the gap between the particles
can be computed analytically~\cite{RMS14,RMS15,FLRC2023}. Another interesting spatial object is not
the instantaneous position of the rightmost particle, but the maximal spatial displacement
of the whole history of the process during the interval $[0,t]$, i.e., the maximal spatial displacement
{\em up to} time $t$. This maximal spatial extent of the BBM in all three phases (supercritical, critical, and
subcritical) have been studied analytically in one dimension~\cite{Mckean75,Bramson78,SF79,RMS15_2}.
In two dimensions, the convex hull of the BBM process up to time $t$ have also been studied
analytically and numerically in the context of epidemic spreads~\cite{DMRZ13}.

\begin{figure}
\includegraphics[width=0.8\textwidth]{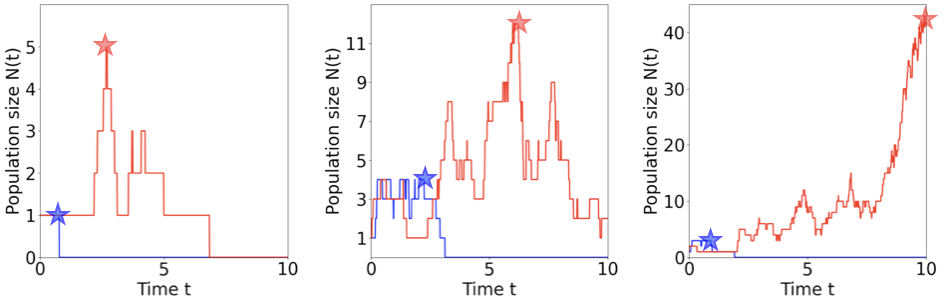}
\caption{Time evolution of the process $N(t)$ for the branching process over the interval $t \in (0, 10)$, shown in 
three regimes: subcritical (left, $a = 1$, $b = 0.8$), critical (middle, $a = b = 1$), and supercritical (right, $a = 
1$, $b = 1.5$). In each panel, two independent realizations are shown (red and blue curves), with their respective 
maximum population sizes $M$ marked by stars. Subcritical case: $M = 5$ (red) and $M = 1$ (blue). Critical case: $M = 
12$ (red) and $M = 4$ (blue). Supercritical case: $M = 43$ (red) and $M = 3$ (blue).}
\label{fig.rw_trajectory}
\end{figure}

While the statistics of the spatial outliers in BBM have thus been studied extensively, in this paper we are
not interested in spatial extents, but rather on the extreme value questions associated
with the other random variable, namely the population size $N(t)$. 
More precisely, consider $N(t)$ as a stochastic process itself (so, space is totally irrelevant
here) that starts at $N(0)=1$ and let $M(t)$ denote the maximal population size {\em up to} time $t$, i.e.,
\begin{equation}
M(t)= \max_{0\le \tau\le t}\, \left[ N(\tau)\right]\, .
\label{max.1}
\end{equation}
Clearly, $M(t)$ is a random variable and we are interested in its 
distribution $Q(L,t)= {\rm Prob.}[M(t)=L]$. This is a natural question in several
contexts mentioned above. For example, in the context of a baceterial colony growth
$Q(L,t)$ provides the statistics of the maximal colony size reached up to time $t$.
In the context of the infection process, $Q(L,t)$ denotes the distribution
of the maximal population of infected people up to time $t$. 
The goal of this
paper is to compute this extreme value distribution $Q(L,t)$ exactly in
all three phases: subcritical ($b<a$), critical ($b=a$) and supercritical ($b<a$).
Typical trajectories of the evolution of the population size $N(t)$ vs. $t$ is shown in Fig. (\ref{fig.rw_trajectory})
in the three phases: subcritical ($b<a$), supercritical ($b>a)$ and critical ($b=a$) respectively and for
each trajectory the corresponding global maximum $M$ is marked as a star.

Obtaining the distribution of the maximum $M(t)$ in \eqref{max.1} for the process $N(t)$ is
nontrivial for the following reason. The statistics of the maximum of a set of $K$ random variables 
$\{z_1,z_2,\ldots, z_K\}$ is well understood when the variables $z_i$'s are uncorrelated and say,
each is distributed via the same probability distribution function (PDF) $f(z)$, i.e., they
are independent and identically distributed (i.i.d) random variables~\cite{Gumbel_book}.
In this case, the distribution of their maximum, in the limit of large $K$, follows
one of the three classical PDF's known as Gumbel, Fr\'echet and Weibull, depending on
the large $z$ behavior of $f(z)$. However, the problem is much harder when
the variables $z_i$'s are strongly correlated, i.e.,  the correlation size is comparable or bigger than 
$K$~\cite{EVS_review, EVS_book}.
There are only a handful examples of such strongly correlated systems for which the distribution of the maximum
can be computed exactly~\cite{EVS_review,EVS_book}. One such simple example being a random walk
of $K$ steps, or a Brownian motion of duration $t$ for which several exact results 
are known with a variety of applications~\cite{EVS_book,
Pollaczek1952,Spitzer1956,CM2005,KM2005,MRKY2008,RMC2009,MCR2010,M2010,SL2010,FM2012,SM12,MMS2013,
CBM2015,GLM2017,MMS2017,MMS19,MMS20,DMS21,DMS23}  
In our case, the underlying process $N(t)$,
whose maximum $M(t)$ up to time $t$ is our object of interest, is an example of a strongly correlated process since
$N(t)$ at two different times are strongly correlated even when the overall duration $t$ becomes large.
Hence, computing exactly the distribution of the maximum of $N(t)$ is also mathematically
challenging. Our exact result for $Q(L,t)$  presented in this paper thus provides another example
of a solvable extreme value problem for a strongly correlated time series or stoachstic process. 

Let us remark that the branching process without spatial diffusion has been studied extensively
in the probability and statistics literature for many decades, mostly in the context of the discrete-time
Galton-Watson trees where at each generation a random number of offsprings are generated~\cite{Harris63}.
Observables such as the population size, as well as the extreme value statistics for this general Galton-Watson
process have been studied extensively and many rigorous results have been derived using probabilitic methods.
The literature is huge (see e.g., the book by T. Harris~\cite{Harris63} and also Refs.~\cite{Lindvall76,Pakes98}
for extreme value questions). However, these probabilistic methods are not always easy to follow by
non-specialists and also often it is not easy to extract explicit results from
this huge literature. The purpose of this article is not to review this extenstive probability literature on general
branching process, but rather focus on a simple version of the branching process that occurs in continuous time
as described above. The advantage of using the continuous time version is that the branching process
can then be mapped exactly to a simple random walk problem on a positive semi-infinite lattice (as demonstrated
later in detail in Section~\ref{random_walk}) where the population size $N(t)$ represents
the position of the random walker on the lattice at time $t$. The associated
hopping probability out of a site labelled $n$ 
is proportaional to $n$. Thus the random walker gets more and more `active'
or `agitated' as it goes farther and further away from the origin. We will call this
effective random walk problem as `agiatated random walk' (in short ARW). The goal of this
paper is to show that
explicit analysis of several observables of the continous time branching process, 
in particular the extreme value statistics (i.e., the statistics of $M(t)$), can
be performed relatively easily by standard pedagogical statistical physics tools using this ARW representation.

It is useful to summarize our main results for the distribution of maximum $Q(L,t)={\rm Prob.}[M(t)=L]$ 
in the three phases in this continuous time branching process. We obtain exact expression for the Laplace
transform of $Q(L,t)$ with respect to $t$, and from this we extract the large $t$ asymptotic behaviors
of $Q(L,t)$:

\begin{itemize}

\item {\bf subcritical phase ($b<a$).} In this case, we show that in the large time limit $t\to \infty$, the
distribution $Q(L, t\to \infty)$ becomes stationary (independent of time) and this stationary
distribution decays exponentially with increasing $L$. This is not surprising, because in the
subcritical phase $b<a$, where death dominates over reproduction, the population eventually becomes extinct
and hence, its maximal population size up to time $t$ becomes independent of time 
as $t\to \infty$.

\item {\bf critical phase ($b=a$).} In this case the distribution $Q(L,t)$ also approaches a stationary
form as $t\to \infty$, but this stationary $Q(L, t\to \infty)$ has a power law tail $\sim L^{-2}$ for large
$L$. For finite but large $t$, we show that $Q(L,t)$ exhibits a scaling form
\begin{equation}
Q(L,t) \approx \frac{1}{L^2}\, f_c\left(\frac{L}{a\, t}\right)\, ,
\label{crit_scaling.1}
\end{equation}
where we compute the scaling function $f_c(z)$ exactly (see Eq. (\ref{exact_fc}) and Fig. (\ref{fig.maximum_critical})).
We show that $f_c(z)\to 1$ as $z\to 0$ and decays as $\approx \sqrt{2}\, z^2\, e^{-z}$ as $z\to \infty$.
Interestingly, the scaling function $f_c(z)$ is nonmonotonic with increasing $z$ with a peak at $z=z^*$.
All moments of the maximum $M(t)$ can also be computed. In particle, the average maximum $\langle M(t)\rangle$
grows very slowly as $\approx \ln (t)$ for large $t$. So, even though the moments grow with time $t$,
the PDF actually approaches a stationary form for $L\ll t$. Essentially the tail of $Q(L,t)$ where $L\sim O(t)$
still has time dependence but the location of this dynamic regime (front region) moves ballistically out 
with a constant speed, leaving behind a stationary PDF.

\item {\bf supercritical phase ($b>a$).} In this case, a completely different scenario emerges for the
PDF $Q(L,t)$ for large $t$. It turns out to have two additive pieces
\begin{equation}
Q(L, t)\approx (-\ln c)\, (1-c)\, c^L \theta\left(
e^{(b-a)t}-L\right) + (1-c)\, \delta\left(L- e^{(b-a)t}\right)\, . 
\label{super_dist.1}
\end{equation}
where $c=a/b<1$ and $\theta(z)$ is the Heaviside theta function with $\theta(z)=1$ for $z>0$ and $\theta(z)=0$ for $z<0$.
The first piece in \eqref{super_dist.1} becomes staionary, i.e., independent of time as $t\to \infty$.
We call this a `fluid' part.
However, the integral of this fluid part, i.e., the total weight carried by it is $c=a/b<1$. 
The remaining probability weight $(1-c)$ is actually carried by the second piece which
approaches a delta function with amplitude $(1-c)$, centered at
$L\approx e^{(b-a)\,t}$. This delta peak at the end of the first piece 
of the PDF is akin to a `condensate' that carries
the additional probability weight $(1-c)$. Physically, this corresponds to trajectories of $N(t)$ that
approach infinity as $t\to \infty$. Indeed, in the supercritical phase, the population survives
eventually (i.e., does not become extinct as $t\to \infty$) with a finite probability $(1-c)$.

\end{itemize}

The rest of the paper is organized as follows. Section \ref{Model} is devoted to the derivation of the population size 
distribution $P(n,t)$ at time $t$ by two rather different methods, respectively in Section \ref{bfp} and 
\ref{random_walk}. The first method involves a backward Master equation that happens to be nonlinear, yet exactly 
solvable.  We will see that this method, however, is not well suited to compute the extremal distribution $Q(L,t)$ of the 
maximum $M(t)$ up to time $t$. Hence, we provide an alternative method of deriving the population size distribution 
$P(n,t)$ at time $t$ in Section \ref{random_walk}. This second method involves, as mentioned above, an exact mapping to an 
auxiliary one dimensional random walk problem on a positive semi-infinite lattice where the hopping probability out of a 
lattice site $n$ actually depends linearly on $n$, i.e., the walker moves faster as it goes away from the origin. We will 
see later that this second method can be suitably adapted to compute the extremal distribution $Q(L,t)$. Section 
\ref{cont_limit} provides a continuous space limit of the auxiliary random walk problem, where we show that the associated 
Fokker-Planck equation in the subcritical phase corresponds to a special case of the well known Feller process in 
probability theory. In Section \ref{max_pop} we provide the exact derivation of the extremal size distribution $Q(L,t)$ up 
to time $t$ using the auxiliary random walk approach. We also present the numerical simulation results that are in 
excellent agreement with our analytical predictions. Finally, we conclude in Section \ref{conclu} with a summary and 
outlook. Some details of the computations are provided in the two appendices.

\section{Population size distribution of a branching process with death} 
\label{Model}

The process starts with a single bacteria at $t=0$. Let $N(t)$
denote the population size of the bacterial colony at time $t$. Clearly, $N(t)$ is
a random variable that fluctuates from one realization of the process to another. We are interested
in computing the size distribution $P(n,t)={\rm Prob.}[N(t)=n]$, i.e., 
the probability that the colony has size $n$ at time $t$. As emphasised earlier, the spatial diffusion
plays no role on the population size distribution and hence, the dynamics of $P(n,t)$ is controlled
by only two parameters: the branching rate $b$ and the death rate $a$.  
There are two alternative methods to derive $P(n,t)$ exactly: the first by employing a backward Fokker-Planck
approach in Section \ref{bfp} and the second by mapping the process to an auxiliary lattice
random walk problem that can be subsequently solved exactly as illustrated in Section \ref{random_walk}.
Finally in Section \ref{cont_limit} we discuss the continuous space limit of the lattice random walk problem and show that for $b<a$ it reduces
to a special case of the Feller process.

\subsection{Size distribution via a backward Fokker-Planck approach}
\label{bfp}

This derivation was presented in Ref.~\cite{RMS14}. Here we reproduce 
it for the sake of completeness. In the ``backward" method, one considers
the time interval $[0, t+\Delta t]$ and splits it into two subintervals:
$[0,\Delta t]$ and $[\Delta t, t+\Delta t]$. There are three possibilities
in the first time interval $[0,\Delta t]$: (i) with probability $b\Delta t$,
the initial particle at $t=0$ splits into two daughter particles, (ii) with
probability $a \Delta t$ the initial particle dies and (iii) with probability
$1-(1+b) \Delta t$, the initial particle does nothing. Let $P(n,t+\Delta t)$
denote the probability of having $n$ particles at time $t+\Delta t$, starting
from a single particle at $t=0$.
Then considering the three events that happen in $[0,\Delta t]$ and
the subsequent evolution over the second interval $[\Delta t, t+\Delta t]$, 
we can write the evolution equation for $P(n,t)$
\begin{equation}
P(n, t+ \Delta t)= \left [1- (a+b)\, \Delta t\right]\, P(n,t)+
b\, \Delta t\, \sum_{n_1, n_2=0}^{\infty} P(n_1,t)\, P(n_2,t)\, \delta_{n+1+n_2,n}+
a\, \Delta t\, \delta_{n,0}\, .
\label{bfp.1}
\end{equation}
Here $\delta_{m,n}$ is the Kronecker delta function which is $1$ if $m=n$
and is zero otherwise.
The first term in Eq. (\ref{bfp.1}) corresponds to the event (iii) when nothing happens in 
$[0,\Delta t]$ and hence the subsequent evolution from $\Delta t$ to 
$t+\Delta t$ of time interval $t$ must be $P(n,t)$. The second term 
corresponds to the branching event (ii) where two independent copies of 
the process starts at $\Delta t$ each with a single particle. The total 
number of particles in the colony at time $t+\Delta t$ is the sum of the 
population sizes from each of these daughters. Similarly, the third term 
corresponds to the event (iii) when the initial particle dies, and hence 
contribute only to the event of zero population size $n=0$ at time 
$t+\Delta t$. This method is called `backward', since we write down
the recursion equation by considering events in the first interval $[0, \Delta t]$ at the begining of the history of evolution. 
Taking the limit $\Delta t\to 0$, gives the {\em nonlinea}r evolution equation
\begin{equation}
\partial_t P(n,t)= - (a+b)\, P(n,t) + b\, \sum_{n_1,n_2=0}^{\infty} P(n_1,t) 
P(n_2,t)\, \delta_{n_1+n_2,n} + a\, \delta_{n,0}\, , \quad n\ge 0
\label{bfp.2}
\end{equation}
This equation starts from the initial condition
\begin{equation}
P(n,0)= \delta_{n,1} \, .
\label{bfp_init.1}
\end{equation}

\begin{figure}
\includegraphics[width=0.8\textwidth]{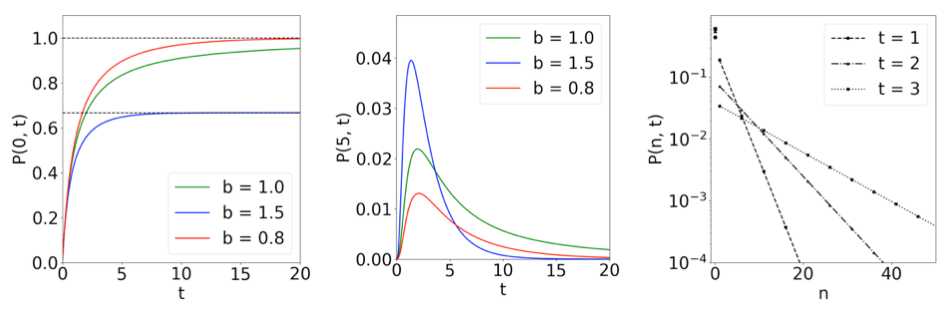}
\caption{
Study of \( P(n,t) \) from Eq.~(\ref{bfp_sol.2}) and Eq.~(\ref{bfp_crit.1}) with \( a = 1 \).
Left: \( P(n=0,t) \) vs.\ \( t \); this probability saturates to 1 in the subcritical and critical cases, while it saturates to \( a/b \) in the supercritical case.
Middle: \( P(n=5,t) \) vs.\ \( t \); the probability displays a non-monotonic behavior and eventually decays to zero at large times.
Right: \( P(n,t) \) vs.\ \( n \) for \( b = 1.5 \); the exponential tail becomes broader as time increases.
}
\label{fig.Pnt}
\end{figure}

To solve Eqs. (\ref{bfp.2}), we introduce the generating function
\begin{equation}
\hat{P}(z,t)= \sum_{n=0}^{\infty} P(n,t)\, z^n \, .
\label{bfp_genf.1}
\end{equation}
Multiplying both sides of Eq. (\ref{bfp.2}) by $z^n$ and summing 
over all $n\ge 0$, we get a first order differential equation
\begin{equation}
\partial_t \hat{P}(z,t)= [1- \hat{P}(z,t)]\,[a- b\, \hat{P}(z,t)]\, ,
\label{bfp_genf.2}
\end{equation}
starting from the initial condition 
\begin{equation}
\hat{P}(z,0)=z \, ,
\label{bfp_genf_init}
\end{equation}
which simply follows from Eq. (\ref{bfp_init.1}). Eq. (\ref{bfp_genf.2})
can be trivially solved, giving
\begin{equation}
\hat{P}(z,t)= \frac{ (a-bz)\, e^{(a-b)t}-a\, (1-z)}
{(a-bz) e^{(a-b)t}- b(1-z)}
\label{bfp_sol.1}
\end{equation} 
Clearly, for $z=1$, we get $\hat{P}(1,t)= \sum_{n=0}^{\infty} P(n,t)=1$, ensuring
the normalization.
Expanding in powers of $z$, one gets the exact distribution of the
population size at any time $t$
\begin{eqnarray}
P(n,t) =
\begin{cases}
&\frac{(b-a)^2 e^{(a-b)t}}{\left[b-a\, e^{(a-b)t}\right]^2}\,
\left[ \frac{b(1- e^{(a-b)t}}{b-a\, e^{(a-b)t}}\right]^{n-1}\, 
\quad {\rm for}\quad n\ge 1  \\
& \\
& \frac{ a\, \left(1- e^{(a-b)t}\right)}{b-a\, e^{(a-b)\, t}}\, 
\quad\quad\quad\quad\quad\quad\quad\quad\,\, {\rm for}\quad n=0\, . 
\end{cases}
\label{bfp_sol.2}
\end{eqnarray}
For the special case $b=a$, by taking the limit $b\to a$ in 
Eq. (\ref{bfp_sol.2}) one gets
\begin{eqnarray}
P(n,t)\Big|_{b=a}=
\begin{cases}
& \frac{1}{(1+a\, t)^2}\, \left[ \frac{at}{1+at}\right]^{n-1} 
\quad {\rm for}\quad n\ge 1 \\
& \\
& \frac{at}{1+at} \, , \quad\quad\quad\quad {\rm for} \quad n=0\, .
\end{cases}
\label{bfp_crit.1}
\end{eqnarray} 
The exact results in Eqs.~\eqref{bfp_sol.2} and \eqref{bfp_crit.1} are shown in Fig.~\ref{fig.Pnt} and are in excellent agreement with numerical simulations.

Note that the average population size $\langle N(t)\rangle$ is given by
\begin{equation}
\langle N(t)\rangle= \sum_{n=1}^{\infty} n\, P(n,t)= 
\partial_z \hat{P}(z,t)\Big|_{z=1}= e^{(b-a)\, t} \, ,
\label{avg_size.1}
\end{equation}
where we used Eq. (\ref{bfp_sol.1}). Hence, for $b>a$, the average
population size grows exponentially with time: this is called 
the `supercritical' phase ($b>a$). In contrast, for $b<a$, the average size vanishes exponentially
for large $t$ leading to an extinction of the bacterail colonly: this
is the `subcritical' phase $b<a$. Exactly for $b=a$, the average size
$\langle N(t)\rangle=1$ does not change with time from its initial value $1$.
This is called the `critical' phase ($b=a$).

From Eq. (\ref{bfp_sol.2}), one can easily work out the asymptotic
behavior of the size distribution at late times $t$. One obtains
the following behaviors.

\begin{itemize}

\item {\bf Subcritical} phase $b<a$. In this case, by taking 
$t\to \infty$ limit in Eq. (\ref{bfp_sol.2}), it follows
that the size distribution for $n\ge 1$ decays
exponentially at late times
\begin{equation}
P(n, t) \approx \left(1- \frac{b}{a}\right)^2\, 
e^{-(a-b) t}\, \left(\frac{b}{a}\right)^{n-1}\,
\label{pnt_subcrit.1}
\end{equation}
while
\begin{equation}
P(0,t\to \infty)\approx 1\, .
\label{pot_subcrit.1}
\end{equation}
Thus, in this subcritical phase, the colony 
becomes extinct exponentially fast
with time. 

\item {\bf Critical} phase $b=a$. In this case, from Eq. (\ref{bfp_crit.1}),
one gets for large $n$
and large $t$
\begin{equation}
P(n,t) \approx \frac{1}{a\, t}\, \frac{1}{n_{\rm c}(t)}\, 
e^{-n/n_{\rm c}(t)}\, ,
\label{c_dist.1}
\end{equation}
where
\begin{equation}
n_{\rm c}(t) \approx a\, t\, .
\label{nc.1}
\end{equation}
Thus in this case, the size distribution again has an exponential tail
over a characteristic size $n_{\rm c}(t)\sim a t$ that grows much slower,
as linearly, with increasing $t$, in contrast to the exponential growth
in the supercritical phase. In this case, it follows from
Eq. (\ref{bfp_crit.1}) that at late times
\begin{equation}
P(0, t\to \infty)=1\, .
\label{pot_crit.1}
\end{equation}
Thus, in the critical case, with probability $1$ the pupulation 
eventually becomes extinct.

\item {\bf Supercritical} phase $b>a$. In this case, one can express
$P(n,t)$ in Eq. (\ref{bfp_sol.2}), for large $n$ and large $t$, in the form
\begin{equation}
P(n, t) \approx \left(1-\frac{a}{b}\right)\, \frac{1}{n_{\rm sc}(t)}\,
e^{- n/n_{\rm sc}(t)}\, ,
\label{sc_dist.1}
\end{equation}
where 
\begin{equation}
n_{\rm sc}(t) = \frac{b}{(b-a)}\, e^{(b-a) t}\, .
\label{nsc.1}
\end{equation}
Thus the size distribution has an exponential tail with a characteristic
size $n_{\rm sc}(t)$ in Eq. (\ref {nsc.1}) that grows exponentially with time.
In contrast, for $n=0$ one gets at late times
\begin{equation}
P(0,t\to \infty)= \frac{a}{b}
\label{p0t.1}
\end{equation}
In fact, integrating Eq. (\ref{sc_dist.1}) over all $n\ge 1$, one
gets $\sum_{n\ge 1} P(n,t) \approx 1- a/b$, which exactly
complements $P(0, t\to \infty)=a/b$. 
Thus, in the supercritical phase $b>a$, 
the population becomes extinct
with a finite probability $a/b<1$, while with the complementary probability
$(1-a/b)$ the size grows exponentially.

\end{itemize}

{\bf Remark:} Let us make a side remark which will be useful later.
In the supercritical phase ($b>a$), the exact size distribution 
$P(n,t)$ in Eq. (\ref{bfp_sol.2}), as a function of $t$ for fixed large $n$,
can be expressed as
\begin{eqnarray}
\label{gumbel.1}
P(n\ge 1, t) &\approx & 
\left(1-\frac{a}{b}\right)^2\, e^{-(b-a) t}\, 
\left[1- \frac{b-a}{b}\, e^{-(b-a)t}\right]^{n-1} \nonumber \\
& \approx & \left(1-\frac{a}{b}\right)^2\,
\exp\left[- (b-a) t- n \frac{(b-a)}{b} \, e^{-(b-a) t} \right] \nonumber \\
& \approx & \left(1-\frac{a}{b}\right)^2\,
\exp\left[- (b-a)\, t - 
e^{- (b-a)\, \left(t- \frac{\ln\left(n\, (b-a)/b\right)}{(b-a)}\right) }
\right]\, .
\end{eqnarray}
Thus, as a function of $t$ for fixed but large $n$, the distribution
$P(n,t)$ has a Gumbel form with a peak at $t=t^*\approx \ln (n)/(b-a)$.   

\subsection{An alternative random walk representation}
\label{random_walk}

\begin{figure}
\includegraphics[width=0.8\textwidth]{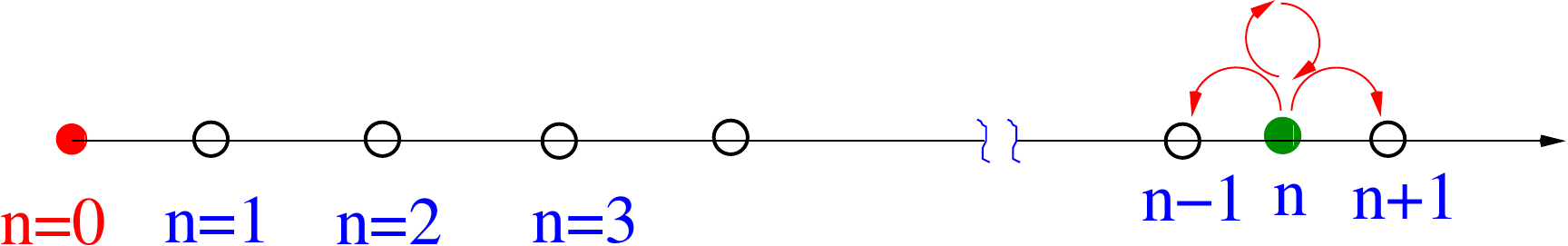}
\caption{A random walker (green) moving on a semi-infinite lattice
whose sites are labelled as $n=0,1,2,3\ldots$. The position of the walker $n$
represents the population size. The walker moves in continuous time with
a site dependent transition rates. In a small time $\Delta t$,
the walker at site $n$ hops to the right neighbour $n+1$ with probability
$b\,n\, \Delta t$,
to the left neighbour $n-1$ with probability $a\, n\, \Delta t$ and 
stays at site $n$
with probability $1- (a+b)\, n\, \Delta t$. The site $n=0$ (shown in red)
is an absorber, i.e., the dynamics stops once the walker reaches the 
site $n=0$.}
\label{fig.rwr1}
\end{figure}

The model of the branching with death, described and solved above, can 
also be solved using an alternative random walk representation. This 
random walk representation turns out to be useful to compute some other 
observables in the model, such as the maximum population size up to time 
$t$, as we will demonstrate later. To derive this alternative representation,
suppose we have $n$ bacteria at time $t$, i.e., $N(t)=n$. In a small time $\Delta t$, each
of these $n$ bacteria attempts to split into two daughters independently.
The probability that $m$ out of $n$ will split is
clearly given by ${n \choose m}(b \Delta t)^m (1- b\, \Delta t)^{n-m}$. 
Hence to leading order in $\Delta t$ as $\Delta t\to 0$, only one out of $n$
will split with probability $ n\, b\, \Delta t$. Thus with this probability
the population size will increase by $1$, i.e., $N(t)\to N(t)+1$. 
Similarly, to leading order
in $\Delta t$, only
$1$ out of $n$ will die with probability $n \, a \Delta t$, i.e.,
$N(t)\to N(t)-1$. Consequently,
the random variable $N(t)$ has the following continuous time dynamics
\begin{eqnarray}
N(t+\Delta t) = \begin{cases}
& N(t)+1 \quad {\rm with\,\, prob.}\quad b\, N(t)\, \Delta t  \\
& \\
& N(t)-1 \quad {\rm with\,\, prob.}\quad a\, N(t)\, \Delta t  \\
& \\
& N(t)  \quad {\rm with\,\, prob.}\quad 1- (a+b)\, N(t)\, \Delta t \, 
\end{cases}
\label{rw_rates.1}
\end{eqnarray}
These transition rates hold for all $n\ge 1$. Note, however, that the
state $n=0$ is special. It is an `absorbing' state. Once the system
reaches $n=0$, it just stays there with probability $1$. Thus, one
can represent this population size $N(t)$ as the position of a random walker on a semi-infinite lattice
whose sites are labelled by $n=0,1,2,\ldots$. A random walker, representing
the population size, hops on this lattice starting from the site $n=1$
at $t=0$. From a site labelled $n$, the walker hops to the site $n+1$
with probability $b\, n\, \Delta t$ (corresponding to branching),
hops to the site $n-1$ with probability $a\, n\, \Delta t$ (corresponding to natural
death) and
stays at site $n$ with the complementary probability 
$1- (a+b)\, n\, \Delta t$. The origin $n=0$ is an `absorbing' site
(see Fig. (\ref{fig.rwr1})). The dynamics stops once the walker reaches the site $n=0$ for the first time.

One can then write a forward Fokker-Planck equation for the population size 
distribution $P(n,t)$ which just represents the probability that the random walker is 
at site $n$ at time $t$, starting from site $1$ at $t=0$.
Imagine evolving the time from $t$ to $t+\Delta t$. Then $P(n, t+\Delta t)$
can be written for all $n\ge 1$
\begin{equation}
P(n, t+\Delta t)= a\, (n+1)\, \Delta t P(n+1,t) + b\, (n-1)\, \Delta t\,  P(n-1,t)
+ \left[1- (a+b)\, \Delta t\right]\, P(n,t)\, .
\label{rw.1}
\end{equation}
In order to arrive at $n$ at time $t+\Delta t$,
the walker at time $t$ must be either at site $n+1$ (from which it hops
to site $n$ with prob. $a\, (n+1)\, \Delta t$), or at site $n-1$ (from which 
it hops to site $n$ with probability $b\, (n-1)\, \Delta t$, or
at site $n$ where it stays with probability $1- (a+b)\, n\, \Delta t$--
explaining the three terms on the right hand side (rhs) of Eq. (\ref{rw.1}).
This is called `forward' since we consider the evolution from time $t$ to time $t+\Delta t$. Taking the limit $\Delta t\to 0$ in Eq. (\ref{rw.1}),
we arrive at a {\em linear} equation for $P(n,t)$
\begin{equation}
\partial_t P(n,t)= - (a+b)\, n \, P(n,t)
+ b\, (n-1)\, P(n-1,t) + a\, (n+1)\, P(n+1,t)\, .
\label{rw.2}
\end{equation}
Note that this equation also holds for $n=0$, provided we interpret
$P(-1,t)=0$. There is no outgoing current from the site $n=0$, but
there is an ingoing current from the 
site $1$. Eq. (\ref{rw.2}) starts from the initial condition
\begin{equation}
P(n,0)= \delta_{n,1}\, .
\label{rw_init.1}
\end{equation}
Note that Eq. (\ref{rw.2}) is linear in $P(n,t)$, while
Eq. (\ref{bfp.2}) is {\em nonlinear} in $P(n,t)$. We will see shortly
that both equations yield the same solution for $P(n,t)$ as it should.
The linear Eq. (\ref{rw_init.1}) has some advantages as we will see later,
however the price we pay is that the transition rates depend on the
site index $n$.

To solve Eq. (\ref{rw.2}), we again define the generating function
as in Eq. (\ref{bfp_genf.1}), namely
\begin{equation}
\hat{P}(z,t)= \sum_{n=0}^{\infty} P(n,t)\, z^n \, .
\label{rw_genf.1}
\end{equation}   
Let us note the identity
\begin{equation}
z\, \partial_z \hat{P}(z,t)= \sum_{n=0}^{\infty} n\, P(n,t)\, z^n\, .
\label{rw_iden.1}
\end{equation}
Multiplying both sides of Eq. (\ref{rw.2}) by $z^n$ and summing over all 
$n\ge 0$ (using $P(-1,t)=0$ and the identity in Eq. (\ref{rw_iden.1})), 
it is easy to show that $\hat{P}(z,t)$ satisfies the first order
equation
\begin{equation}
\partial_t \hat{P}(z,t)= (1-z)\, (a-b\, z)\, \partial_z \hat{P}(z,t)\, ,
\label{rw_evol.1}
\end{equation}
starting from the initial condition
\begin{equation}
\hat{P}(z,t=0)= z\, .
\label{rw_evol_init.1}
\end{equation}
The general solution of a first-order equation of the type (\ref{rw_evol.1})
can be expressed as a travelling front (using the method of characteristics)
\begin{equation}
\hat{P}(z,t) = F\left(t+ f(z)\right)\, ,
\label{rw_sol.1}
\end{equation}
where both $f(z)$ and $F(u)$ are yet to be fixed. Substituting the
anticipated solution (\ref{rw_sol.1}) in Eq. (\ref{rw_evol.1}), one
immediately gets a first-order differential equation for $f(z)$ (while
the function $F(u)$ is still arbitrary)
\begin{equation}
\frac{df}{dz}= \frac{1}{(1-z)(a-b\, z)}\, .
\label{rw_fsol.1}
\end{equation}
The general solution of Eq. (\ref{rw_fsol.1}) is given by
\begin{equation}
f(z)= \frac{1}{a-b}\, \ln \left( \frac{(a-b\, z)}{1-z}\right) +C_0\, ,
\label{rw_fsol.2}
\end{equation}
where $C_0$ is an arbitrary constant. Substituting this solution for $f(z)$
in Eq. (\ref{rw_sol.1}), we then get the general solution in the form
\begin{equation}
\hat{P}(z,t)= F\left(t+ C_0+ \frac{1}{a-b}\, 
\ln \left( \frac{(a-b\, z)}{1-z}\right) \right)\, .
\label{rw_sol.3}
\end{equation}
The important point to note is that this solution is valid for any time $t$,
including $t=0$. Thus the function $F(u)$ can then be fixed from the initial
condition in Eq. (\ref{rw_evol_init.1}). Indeed, setting $t=0$
in Eq. (\ref{rw_sol.3}) and using Eq. (\ref{rw_evol_init.1}) gives
\begin{equation}
F\left(C_0+ \frac{1}{a-b}\,
\ln \left( \frac{(a-b\, z)}{1-z}\right) \right)= z\, .
\label{Fsol.1}
\end{equation}
Setting
\begin{equation}
u = C_0 + \ln \left( \frac{(a-b\, z)}{1-z}\right) \, ,
\label{u_def.1}
\end{equation}
or equivalently
\begin{equation}
z= 1- \frac{a-b}{e^{(a-b)(u-C_0)}-b} \, ,
\label{zvsu.1}
\end{equation}
we obtain from Eq. (\ref{Fsol.1})
\begin{equation}
F(u)=z= 1- \frac{a-b}{e^{(a-b)(u-C_0)}-b}\, .
\label{Fsol.2}
\end{equation}
Finally, substituting this on the rhs of Eq. (\ref{rw_sol.3}) gives
us the desired solution explicitly, which upon simplification, reads
\begin{equation}
\hat{P}(z,t)= \frac{ (a-bz)\, e^{(a-b)t}-a\, (1-z)}
{(a-bz) e^{(a-b)t}- b(1-z)}
\label{rw_sol.4}
\end{equation}
which, of course, coincides with the solution in Eq. (\ref{bfp_sol.1})
obtained via the backward nonlinear equation. Consequently, one obtains
explicitly $P(n,t)$ as in Eq. (\ref{bfp_sol.2}) for all $n$ and all $t$.
Thus, quite remarkably, the size distribution $P(n,t)$ satisfies
simultaneously two very different evolution equations, 
one nonlinear in Eq. (\ref{bfp.2})
and the other linear as in Eq. (\ref{rw.2}), but with the same final solutions.
We will see in the next section that for computing the distribution
of the maximal population size up to a fixed time $t$, the second approach
using the random walk representation is much more convenient. 

\subsection{The continuum limit and the Feller process}
\label{cont_limit}

In our branching process with death in continuous time, the population size 
$n$ is a discrete non-negative integer. For large $n$, one can approximate this 
process by an auxiliary process that takes place in continuous space. To 
arrive at this continuum limit, we first introduce a lattice spacing $\delta$ 
and write
\begin{equation}
x= n\, \delta \, .
\label{Cont.1}
\end{equation}
Under this rescaling the probability distribution $P(n,t)$ of the population 
size $n$ must scale as
\begin{equation}
P(n,t) = \delta\, \tilde{P}(n\delta, t) \,
\label{Cont.2}
\end{equation}
where $\tilde{P}(x,t)$ can be interpreted as the 
probability density of the population in the continuous space
variable $x$.
Note that the scaling in Eq. (\ref{Cont.2}) also ensures the normalization
\begin{equation}
\sum_{n=0}^{\infty} P(n,t) = \int_{0}^{\infty} P(x,t)\, dx =1\, ,
\label{Cont.3}
\end{equation}
upon identifying $dx=\delta$. The continuum limit corresponds to taking the 
limit $\delta\to 0$,
$n\to \infty$, keeping the product $x=n\,\delta$ fixed. 
We then substitute this scaling form (\ref{Cont.2}) in the evolution
equation (\ref{rw.2}), replace $n=x/\delta$ and expand for small $\delta$. 
Keeping terms upto order $\delta$, we get
\begin{equation}
\partial_t \tilde{P}(x,t) = \frac{1}{2} (b+a)\, \delta\, x\, 
\partial_x^2 \tilde{P}(x,t) + (a-b)\, \partial_x \left(x\,
\tilde{P}\right)\, + (a+b)\, \delta\, \partial_x \tilde{P} 
= 
\frac{1}{2} (b+a)\, \delta\,  
\partial_x^2 \left(x\, \tilde{P}\right)
- (b-a) \partial_x \left(x\, \tilde{P}\right)\, .
\label{Cont.4}
\end{equation}
Next we divide both sides of Eq. (\ref{Cont.4}) by $2\,b\,\delta$ and
define a rescaled time,
\begin{equation}
T= 2\, \delta\, b\, t\, .
\label{T_def}
\end{equation} 
In this rescaled time,
Eq. (\ref{Cont.4}) reduces to
\begin{equation}
\partial_T \tilde{P}(x,T) =
\frac{(b+a)}{4b}\, \partial_x^2 \left(x\, 
\tilde{P}\right)
-\frac{(b-a)}{2\,\delta\, b}\, \partial_x \left(x\, \tilde{P}\right)\, .
\label{Cont.5}
\end{equation}
We then see that in order to have a well defined continuum limit 
when $\delta\to 0$, we must also scale
\begin{equation}
b-a= 2\, b\, B\, \delta\, , 
\label{Cont.6}
\end{equation}
where $B$ is a constant.
Thus one must take the branching rate $b$ and annihilation rate $a$ very 
close to each other, i.e.,
$b-a\to 0$, $\delta\to 0$ while keeping the ratio 
$B= (b-a)/(2b\delta)$ fixed. In this limit,
we then have a well defined Fokker-Planck equation in the continuous 
semi-infinite line $x> 0$ that reads
\begin{equation}
\partial_T \tilde{P}(x,T)= \frac{1}{2}\, 
\partial_x^2 \left(x\, \tilde{P}\right)
- B\, \partial_x \left(x\, \tilde{P}\right)\, , \quad {\rm where} \quad 
B= \frac{(b-a)}{2\,b\,\delta}\, .
\label{Cont.7}
\end{equation}
Note that $x=0$ is a special sink site (as in the lattice model). Once the particle reaches $x=0$, the
dynamics is over and the process stops.

The Fokker-Planck equation (\ref{Cont.7}) describes a process on the 
continuous semi-infinite line $x\ge 0$ with a space dependent drift and 
diffusion. Indeed, the Langevin equation that gives rise to this Fokker-Planck 
equation (\ref{Cont.7}) can be written explicitly as
\begin{equation}
\frac{dx}{dT}= B\, x + \sqrt{x}\, \eta(T) \,
\label{Cont.9}
\end{equation}
where $\eta(T)$ is a Gaussian white noise with zero mean and a 
delta correlator,
$\langle \eta(T)\eta(T')\rangle= \delta(T-T')$. This Langevin Eq. (\ref{Cont.9}) has a multiplicative noise
term $\sqrt{2 D(x)}\, \eta(t)$ with $D(x)=x/2$ and it has to be interpreted in the It$\hat{\rm o}$ sense. 
Thus the effective diffusion constant $D(x)=x/2$
vanishes as $x\to 0$.
Identifying the drift term $B\, x= - \partial_x V(x)$,
the Langevin equation (\ref{Cont.9}) corresponds to the motion of an 
overdamped particle in an external harmonic potential
\begin{equation}
V(x)= -\frac{B}{2} x^2\, ,
\label{Cont.10}
\end{equation}
Note that in our problem, the parameter $B= (b-a)/(2b \delta)$ can be both 
positive or negative. In fact, $B<0$ ($b<a$) corresponds to the 
`subcritical' phase while $B>0$ ($b>a$) corresponds to the `supercritical' phase. The 
case $B=0$  
corresponds to the `critical' case $b=a$. One important point about this process is 
that both the drift and the diffusion term vanishes as $x\to 0$, making the origin a sink. 
The process terminates when the particle reaches $x=0$, starting from some initial position $x_0>0$. 

To summarize, the continuum limit makes sense when $n\to \infty$, $t\to \infty$
and $(b-a)\to 0$ such that
\begin{equation}
b-a= 2\,b\,B\,\delta\, , \quad\,   t= \frac{T}{2b\delta}\, , \quad\, n=\frac{x}{\delta}\, 
\label{rescaling.1}
\end{equation}
with $B$, $T$ and $x$ held fixed.
One can now check that the exact solution $P(n,t)$, obtained for discrete $n$
in Eq. (\ref{bfp_sol.2}), does produce an exact solution of
the Fokker-Planck equation (\ref{Cont.7})
in this continuum limit given in Eq. (\ref{rescaling.1}) 
For example, consider
the subritical case $b<a$. Here the exact solution for $P(n,t)$
from Eq. (\ref{bfp_sol.2}) for $n>0$ is given by (with $b<a$)
\begin{equation}
P(n,t)= \frac{(b-a)^2 e^{(a-b)t}}{\left[b-a\, e^{(a-b)t}\right]^2}\,
\left[ \frac{b(1- e^{(a-b)t}}{b-a\, e^{(a-b)t}}\right]^{n-1}\,
\quad {\rm for}\quad n\ge 1  \, .
\label{exact_sol.1}
\end{equation}
By taking the scaling limit as in Eq. (\ref{rescaling.1}) and using
the correspondence in Eq. (\ref{Cont.2}), one gets
\begin{equation}
\tilde{P}(x,T)= C\, \frac{e^{|B|T}}{\left(e^{|B|T}-1\right)^2}\, 
\exp\left[-\frac{2 |B| e^{|B|T}}{\left(e^{|B|T}-1\right)}\, x\right]\, ,
\label{cont_sol.1}
\end{equation}
where $|B|=-B= (a-b)/(2b\delta)>0$ and $C= 4 |B|^2\, \delta$. 
One can now check by direct substitution, that the solution in Eq. (\ref{cont_sol.1})
indeed satisfies the Fokker-Planck equation (\ref{Cont.7}).
The total mass contained in positive semi-infinite axis is then given by (to leading order in $\delta$)
\begin{equation}
\int_0^{\infty} \tilde{P}(x,T)\, dx= \frac{2\,|B|\,\delta}{e^{|B|T}-1}\, .
\label{mass.1}
\end{equation}
This indicates that the total density of particles in $x>0$ decreases monotonically with increasing $T$, demonstrating
a nonzero leakage current to the sink site $x=0$. To check this, we note that exactly at $n=0$
the discrete solution \eqref{bfp_sol.2} reads
\begin{equation}
P(0,t)= \frac{ a\, \left(1- e^{(a-b)t}\right)}{b-a\, e^{(a-b)\, t}} \, .
\label{exact_sol_n0.1}
\end{equation}
Taking the scaling limit \eqref{rescaling.1} of this solution, one gets, to $O(\delta)$ 
\begin{equation}
P(0,T)\approx 1- \frac{2\,|B|\,\delta}{e^{|B|T}-1}\, .
\label{cont_sol_x0.1}
\end{equation}
From \eqref{mass.1} and \eqref{cont_sol_x0.1}, it follows that
\begin{equation}
\int_0^{\infty} \tilde{P}(x,T)\, dx + P(0,T)= 1\, ,
\label{norm.1}
\end{equation}
ensuring that the total probability is conserved at all time $T$ if we take into acount the sink site at $x=0$.

The Langevin process in Eq. (\ref{Cont.9}) is a particular case of the
celebrated Feller process that describes an overdamped particle
in the presence of a linear drift and a space dependent diffusion
constant~\cite{Feller51}
\begin{equation}
\frac{dX}{dT}= -\alpha\, X(T) +\beta + \sqrt{X(T)}\, \eta(T)\, ,
\label{feller.1}
\end{equation}
where $X(T)\ge 0$, $\alpha>0$ and $\beta$ is a constant. The noise
$\eta(T)$ is a Gaussian white noise with zero mean and a correlator
$\langle \eta(T)\eta(T')\rangle= \delta(T-T')$. As before, Eq. (\ref{feller.1})
is interpreted in the It$\hat{\rm o}$ sense. The Feller process (\ref{feller.1})
has been studied extensively and has found numerous applications
across disciplines: from neurobiology~\cite{CR73,CR74} to 
finance~\cite{CIR85,Hull,Heston93,DY2002} and
its first-passage properties have been investigated more 
recently~\cite{MP2012,Somrita2022}. The continuum limit
of our branching process in Eq. (\ref{Cont.9}) in the
subcritical case $(b<a)$ corresponds to Feller process
in Eq. (\ref{feller.1}) with $\alpha= -B>0$ and $\beta=0$.
However, the supercritical phase $b>a$, i.e., $\alpha<0$
is usually not studied in the context of the Feller process.
Thus, the Feller process may provide some asymptotic information
about the population size distribution in the branching process for large $n$ in the subcritical
phase, but the discrete time
random walk problem is much richer. In particular, it allows us to study the supercritical and the critical
regime, thus going beyond the Feller process.

\section{Maximal population size up to a fixed time $t$}
\label{max_pop}

In this section, we study the main object of our interest in the branching process,
namely the maximum population size up to a fixed time $t$, starting from a
single bacteria. More precisely, if $N(\tau)$ represents the
population size at time $\tau$ starting from $N(0)=1$, 
we define the maximal size up to time $t$ as
\begin{equation}
M(t)= \max_{0\le \tau\le t} \left[ N(\tau)\right]\, .
\label{max_def.1}
\end{equation}
Clearly, $M(t)$ is a random variable and in this section we compute exactly its
distribution ${Q(L,t)=\rm Prob.}[M(t)=L]$. We first tried to compute this distribution
using the traditional backward Fokker-Planck approach used for the derivation of the
population size distribution detailed in Section \ref{bfp}. But very quickly, we realized
that this method is not suitable to compute the distribution of the maximum $M(t)$ up to time $t$.
This is because of the following technical reason. 
At any given time $t$, the total population size $N(t)=N_1(t)+N_2(t)$, where $N_1(t)$ and $N_2(t)$
denote the population sizes of the two daughter branches after the first branching event.
Thus one needs to calculate the distribution of the maximum of the sum 
$M(t)= \max_{0\le \tau\le t}\, [N_1(\tau)+N_2(\tau)]$. It is not obvious how to write
a recursion relation that keeps track of the maximum of the sum of two independent
populations. In contrast, it is much easier to compute the distribution
of $M(t)$ using the second random walk approach depicted in Fig. (\ref{fig.rwr1}) and discussed in detail
in Section \ref{random_walk}. The main idea is to study this random walk problem by placing 
an absorbing barrier at some value $L>0$. Then, the survival probability of the walker up to time $t$ starting
initially at $N(0)=1$
(i.e., the probability that the walker stays strictly in the regime $0<n<L$ up to time $t$) is
precisely the probability that the maximum $M(t)$ of the walker up to time $t$ is less than $L$. Thus, this
survival probability gives access to the cumulative distribution of the maximum $M(t)$ and from this,
one can compute the PDF of $M(t)$. The survival probability up to time $t$ can be computed exactly, leading
eventually to an exact derivation of the PDF $Q(L,t)= {\rm Prob.}[M(t)=L]$. This is the strategy we follow
in detail below.

To proceed, we first define for this random walk problem
\begin{equation}
q(n,t|L)= {\rm Prob.}\left[{\rm {the\,\, walker\,\, does\,\, not\,\, reach\,\, 
the\,\, site\,\, L\,\, up\,\, to\,\, time\,\, t, \,\,
starting\,\, at\,\, the\,\, initial\,\, position\,\, n}}\right] \, .
\label{q_def.1}
\end{equation}
Then, $q(n,t|L)$ is also the probability that the maximum $M(t)$ of the
walk up to time $t$, starting at the site $n$, stays below $L$. Hence,
this is just the cumulative distribution of the maximum up to $t$ for the process $N(t)$ starting at
$N(0)=n$. In our case, the walker starts at $n=1$, and hence
\begin{equation}
q(1,t|L)= 
{\rm Prob.}\left[M(t)\le L\Big|{\rm {starting\,\, at\,\, n=1\,\, 
at\,\, t=0}}\right]
\label{q1_cum.1}
\end{equation}
Once we know this cumulative distribution, the probability distribution 
function (PDF) of $M(t)$, starting from $N(0)=1$, can be obtained from it simply by
\begin{equation}
Q(L,t)= {\rm Prob.}\left[M(t)=L\right]= p(1,t|L)= q(1,t|L)-q(1,t|L-1)\approx 
\partial_L q(1,t|L) \, ,
\label{p1_pdf.1}
\end{equation}
where the last step holds for large $L$ where one can replace the discrete $L$
by its continuum analogue. Our next goal is to write down the evolution equation
for $q(n,t|L)$ for general $n$. From the solution for $q(n,t|L)$, we
will then be able to compute the cumulative distribution of the maximum
$q(1,t|L)$ and then derive the PDF of the maximum $M(t)$ using 
Eq. (\ref{p1_pdf.1}). 

To write the evolution equation for $q(n,t|L)$, we first make a shorthand
notation $q(n,t|L)\equiv q(n,t)$ for simplicity. The dependence on $L$ is
implicit and will be restored later. It is easy to write down the
backward evolution equation for $q(n,t)$ as
\begin{equation}
\partial_t q(n,t)= b\, n\, q(n+1,t)+ a\, n\, q(n-1,t)- (a+b)\, n\, q(n,t)\, ,
\label{q_bfp.1}
\end{equation}
valid for all $1\le n\le L-1$, and satisfying the boundary conditions
\begin{equation}
q(0,t)=1\, , \quad {\rm and}\quad q(L,t)=0\, ,
\label{q_bc.1}
\end{equation}
and the initial condition
\begin{equation}
q(n,0)=1\, , \quad {\rm for}\,\, 0\le n\le L-1 \, .
\label{q_init.1}
\end{equation}
Eq. (\ref{q_bfp.1}) is easy to understand. Note that $q(n,t)$ is the `survival'
probability that the walker, starting at $n$ at $t=0$, does not reach $L$
up to time $t$. In the first time interval $\Delta t$, the walker
moves from $n$ to $n+1$ with probability $b\,n\, \Delta t$, to site
$n-1$ with probability $a\,n\, \Delta t$ and with the rest of the
probability $1- (a+b)\, n\, \Delta t$ it stays at site $n$. Writing
the three events explicitly, and taking the limit $\Delta t\to 0$
gives Eq. (\ref{q_bfp.1}). The boundary conditions in Eq. (\ref{q_bc.1})
follow from the fact that if the walker starts at $n=0$ it stays there
forever and hence never reaches the site $L$, explaining $q(0,t)=1$.
Similarly, if the walker starts at $L$, clearly its survival probability
(i.e., the probability of not reaching $L$) is zero. The initial
condition in Eq. (\ref{q_init.1}) also follows immediately, since
if the particle starts at any site $0\le n<L$, it can not reach the site $L$
at time $t=0$, hence its survival probability is $1$.

To solve Eq. (\ref{q_bfp.1}), it is convenient to define the
Laplace transform with respect to $t$
\begin{equation}
\tilde{q}(n,s)= \int_0^{\infty} q(n,t)\, e^{-s\,t}\, dt \, .
\label{q_lt.1}
\end{equation}
Taking Laplace transform of Eq. (\ref{q_bfp.1}) with respect to $t$
and using the initial condition $q(n,0)=1$ for all
$0\le n\le L-1$ we get
\begin{equation}
b\, n\, \tilde{q}(n+1,s)+ a\, n\, \tilde{q}(n-1,s) 
- (a+b)\, n\, \tilde{q}(n,s)= -1 + s\, \tilde{q}(n,s)\, .
\label{q_lt.2}
\end{equation}
with the boundary conditions (from Eqs. (\ref{q_bc.1}))
\begin{equation}
\tilde{q}(0,s)= \frac{1}{s}\, , \quad {\rm and}\quad \tilde{q}(L,s)=0\, .
\label{q_bc.2}
\end{equation}
It is further convenient to make the shift
\begin{equation}
\tilde{q}(n,s)= \frac{1}{s} + \tilde{r}(n,s)
\label{shift.1}
\end{equation}
in Eq. (\ref{q_lt.2}) that makes the equation homogeneous, i.e.,
\begin{equation}
b\, n\, \tilde{r}(n+1,s)+ a\, n\, \tilde{r}(n-1,s)
- \left[(a+b)\, n+s\right]\, \tilde{r}(n,s)=0\, , \quad {\rm for}\quad 
1\le n\le L-1\, ,
\label{r_lt.1}
\end{equation}
with the associated boundary conditions
\begin{equation}
\tilde{r}(0,s)=0\, \quad {\rm and}\quad \tilde{r}(L,s)=-\frac{1}{s}\, .
\label{r_bc.1}
\end{equation}

To solve Eq. (\ref{r_lt.1}) over the finite interval $1\le n\le L-1$,
with the boundary conditions (\ref{r_bc.1}), it is actually convenient
to consider an auxiliary problem in which
Eq. (\ref{r_lt.1}) is assumed to hold in the full semi-infinite
line $n\ge 0$, with the two conditions in Eq. (\ref{r_bc.1}). The
solution for $n>L$ in this auxiliary semi-infinite problem
is of no importance to us, but its solution for $n\le L$
will coincide with our finite interval problem. It turns out
that the solution of the auxiliary semi-infinite problem is
much simpler, compared to the finite interval problem. Thus,
the extension to this full semi-infinite lattice is purely 
for mathematical convenience.

To proceed, we define the generating function
\begin{equation}
W(z,s)=\sum_{n=1}^{\infty} \tilde{r}(n,s)\, z^n \, .
\label{r_genf.1}
\end{equation}
Multiplying Eq. (\ref{r_lt.1}) by $z^n$, summing over all $n\ge 1$ and
after some simplifying algebra (and using $\tilde{r}(0,s)=0$), we get a first-order differential equation
\begin{equation}
\left[a\, z^2- (a+b)\,z +b\right]\,\partial_z W(z,s)= 
\left(\frac{b}{z}+s-a\, z\right)\, W(z,s) \, .
\label{wzs.1}
\end{equation}  
This equation can be trivially solved to give
\begin{equation}
W(z,s)= A\, \frac{z}{(1-z)^{1+s/(b-a)}}\, \frac{1}{(b-a\,z)^{1-s/(b-a)}}\, ,
\label{wzs_sol.1}
\end{equation}
where $A$ is an arbitrary constant, yet to be fixed. Inverting 
Eq. (\ref{r_genf.1}) formally using Cauchy's inversion formula and
using Eq. (\ref{wzs_sol.1}) we get
\begin{equation}
\tilde{r}(n,s)= \oint \frac{dz}{2\pi\, i}\, \frac{1}{z^{n+1}}\, W(z,s)
= A\, \oint \frac{dz}{2\pi\, i}\, \frac{1}{z^n}\, \frac{1}{(1-z)^{1+s/(b-a)}\,
(b-a\, z)^{1-s/(b-a)}}\, .
\label{rns_sol.1}
\end{equation}
We can now fix the unknown constant $A$, using the condition
$\tilde{r}(L,s)=-1/s$ in Eq. (\ref{r_bc.1}). This gives
\begin{equation}
A= -\frac{1}{s}\, \frac{1}{ \oint \frac{dz}{2\pi\, i}\, \frac{1}{z^L}\, \frac{1}{(1-z)^{1+s/(b-a)}\,
(b-a\, z)^{1-s/(b-a)}} }\, .
\label{A.1}
\end{equation}
Once we have $\tilde{r}(n,s)$ from Eq. (\ref{rns_sol.1}), we can finally get
our desired $\tilde{q}(n,s)=1/s+\tilde{r}(n,s)$ explicitly as
\begin{equation}
\tilde{q}(n,s)= \frac{1}{s}\left[1- \frac{\tilde{I}_n(s)}{\tilde{I}_L(s)}\right]\, ,
\label{qns_sol.1}
\end{equation}
where 
\begin{equation}
\tilde{I}_n(s)= \oint \frac{dz}{2\pi\, i}\, \frac{1}{z^n}\, \frac{1}{(1-z)^{1+s/(b-a)}\,
(b-a\, z)^{1-s/(b-a)}}\, .
\label{Ins.1}
\end{equation}
In particular, one finds that 
for $n=1$ (i.e., when the walker starts at site $n=1$), the
expression for $\tilde{I}_1(s)$ simplifies 
\begin{equation}
\tilde{I}_1(s)= \frac{1}{b^{1- s/(b-a)}}\, .
\label{I1s.1}
\end{equation}
This follows from the fact that $\tilde{I}_1(s)$ is just the coefficient of $z^0$
in the power series expansion of the factor 
$(1-z)^{-1-s/(b-a)} (b-az)^{-1+s/(b-a)}$. Hence, from Eq. (\ref{qns_sol.1}),
we get, for $n=1$
\begin{equation}
\tilde{q}(1,s)= \int_0^{\infty} q(1,t|L)\, e^{-s\, t}\, dt
= \frac{1}{s}\left[1-\frac{1}{\tilde{J}_L(s)}\right]\, ,
\label{q1s.2}
\end{equation}
with
\begin{eqnarray}
\tilde{J}_L(s) &= &\oint \frac{dz}{2\pi\, i}\, \frac{1}{z^L}\, \frac{1}{(1-z)^{1+s/(b-a)}\,
\left(1-\frac{a\, z}{b}\right)^{1-s/(b-a)}} \\ 
&= & \left(\frac{a}{b}\right)^{L-1}\, 
\frac{\Gamma\left(L-\frac{s}{b-a}\right)}{\Gamma(L)\, 
\Gamma\left(1-\frac{s}{b-a}\right)}\, 
{}_2F_1\left(-L+1,1+\frac{s}{b-a}, -L+1+\frac{s}{b-a}, \frac{b}{a}\right)\, ,
\label{JLs.1}
\end{eqnarray}
where ${}_2F_1(\alpha,\beta,\gamma,z)$ is the Gauss hypergeometric funnction.
The last identity follows from Mathematica. In principle,
the results in Eqs. (\ref{q1s.2}) and (\ref{JLs.1}) are exact
for all $a$ and $b$. However, one has to be a bit careful
for the critical case $a=b$ (the limit $a\to b$ has to be taken
carefully). In fact, in Appendix \ref{A1}, we show that in the critical case $a=b$,
Eq. (\ref{q1s.2}) reduces to
\begin{equation}
\tilde{q}(1,s)\Big|_{b=a}= 
\int_0^{\infty} q(1,t|L)\, e^{-s\, t}\, dt
= \frac{1}{s}\left[1-\frac{1}{L\,\, {}_1F_1\left(1-L,2,-\frac{s}{a}\right)}
\right]\, 
\label{q1s_crit.2}
\end{equation}

From the exact Laplace transform $\tilde{q}(1,s)$ in Eq. (\ref{q1s.2}),
one can extract the late time behavior of the cumulative distribution
$q(1,t|L)$ of the maximum $M(t)$. Below we summarize this asymptotic behavior
of the distribution of $M(t)$ in the three phases. For convenience, we
define the parameter
\begin{equation}
c= \frac{a}{b} \, .
\label{c_def}
\end{equation}
We will consider the subcritical $(c>1)$, critical $(c=1)$ and supercritical
$(c<1)$ phases separately.

\begin{itemize}

\item {\bf Subcritical} phase $c=a/b>1$. To extract the late time behavior
of $q(1,t|L)$, we need to analyse the small $s$ behavior of its Laplace
transform in Eq. (\ref{q1s.2}). We first note that for $c>1$, the
intgral $\tilde{J}_L(s)$ has a finite limit as $s\to 0$. In fact, by setting $s=0$ in
Eq. (\ref{JLs.1}), we get
\begin{equation}
\tilde{J}_L(0)= \oint \frac{dz}{2\pi\, i}\, \frac{1}{z^L}\, 
\frac{1}{(1-z)\, (1-cz)}= \frac{c^L-1}{c-1}\, .
\label{sub_JL0.1}
\end{equation}
The last identity follows simply by expanding $1/[(1-z)(1-cz)]$ in 
powers of $z$ and computing the coefficient of $z^{L-1}$ in this 
expansion. Hence, from Eq. (\ref{q1s.2}), we get
\begin{equation}
\tilde{q}(1,s) \xrightarrow[s\to 0]{} \frac{1}{s}\left[1- \frac{c-1}{c^L-1}\right]\, .
\label{sub_q1s.1}
\end{equation}
Inverting the Laplace transform trivially, it follows that
$q(1,t|L)$ approaches a time-independent value as $t\to \infty$
\begin{equation}
q(1,t|L) \xrightarrow[t\to \infty]{} 1- \frac{c-1}{c^L-1}\, .
\label{sub_q1t.1}
\end{equation}
First note that since $c>1$, we have $q(1,t\to \infty|L)\to 1$ as $L\to \infty$.
Since $q(1,t|L)$ is the probability that $M(t)\le L$, it follows
that the PDF of $M(t)$ is normalized to unity.
Indeed, for large $L$, it follows from Eq. (\ref{p1_pdf.1}), that
the PDF of the maximum $M(t)$ decays exponentially
\begin{equation}
Q(L, t\to \infty)={\rm Prob.}[M(t)=L]=p(1,t\to \infty|L) \approx \frac{(c-1)\,\ln c}{c^{L}}\, .
\label{sub_Mpdf.1}
\end{equation}
This result in Eq.~\eqref{sub_Mpdf.1} is shown in Fig.~\ref{fig.maximum_subcritical} and is in good agreement with numerical simulations for large $t$.

\begin{figure}
\includegraphics[width=0.8\textwidth]{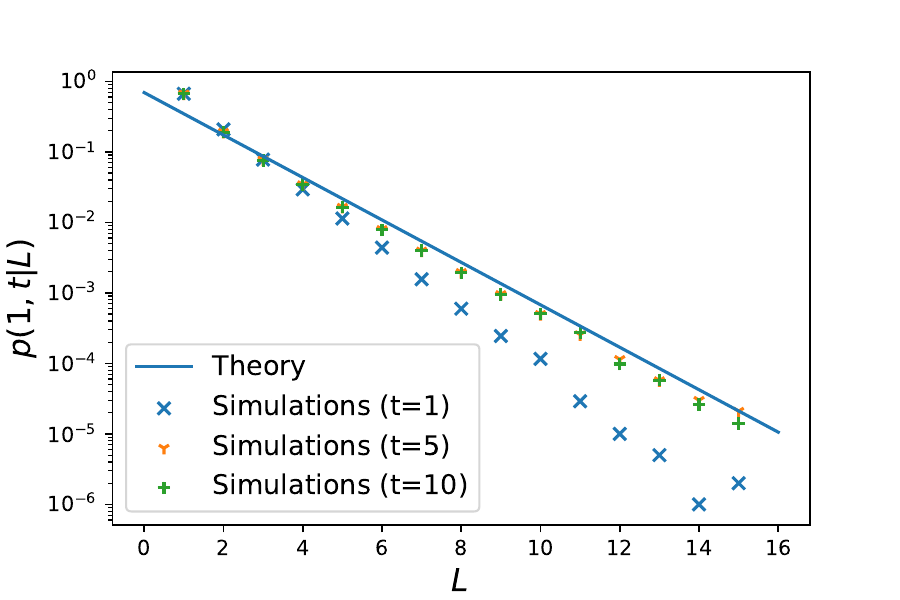}
\caption{Probability distribution $p(1,t|L)$ of the maximal population size $L$ up to time $t$. The continuous blue line corresponds to the asymptotic result in Eq.~\eqref{sub_Mpdf.1}, while the symbols correspond to numerical simulations with $a=2$, $b=1$ and different values of $t$.}
\label{fig.maximum_subcritical}
\end{figure}

\item {\bf Critical} phase $c=a/b=1$. In this case, we analyse
Eq. (\ref{q1s_crit.2}) in the limit $s\to 0$. Using
${}_1F_1(1-L,2,0)=1$, we see that
\begin{equation}
\tilde{q}(1,s) \xrightarrow[s\to 0]{} \frac{1}{s}\left[1-\frac{1}{L}\right]\, .
\label{crit_q1s.1}
\end{equation}
Consequently, inverting the Laplace transform, we see that $q(1,t|L)$ 
approaches a time-independent distribution as $t\to \infty$
\begin{equation}
q(1,t|L) \xrightarrow[t\to \infty]{} 1-\frac{1}{L}\, .
\label{crit_q1t.1}
\end{equation}
Hence the PDF of the maximum, from Eq. (\ref{p1_pdf.1}), decays
as a power law for large $L$
\begin{equation}
Q(L, t\to \infty)= p(1,t\to \infty|L)= {\rm Prob.}[M(t)=L] \approx \frac{1}{L^2}\, .
\label{crit_Mpdf.1}
\end{equation}
Once again, this PDF is normalized to unity, $\int_1^L dL/L^2=1$.
Actually, for finite but large $t$, one can show that the PDF
$p(1,t|L)$ as a function of $L$, gets cut-off at a scale $L_c(t)\sim a\, t$.
More precisely, for large but finite $t$, the PDF $p(1,t|L)$ approaches
a scaling form
\begin{equation}
p(1,t|L) \approx \frac{1}{L^2}\, f_c\left(\frac{L}{a\,t}\right)\, ,
\label{crit_larget.1}
\end{equation}
where the scaling function $f_c(z)$ reads (see Appendix B)
\begin{equation}
f_c(z)= 1+\sum_{n=1}^{\infty}\frac{1}{{}_0F_1(1,-p_n^2/4)}\left(1-\frac{p_n^2}{4z}\right)e^{-p_n^2/(4z)}\,,
\label{exact_fc}
\end{equation}
where ${}_0F_1(1,z)$ is the confluent hypergeometric function and  
$p_n $ is the $n$-th negative root of $J_1(p)=0$ (with $0>p_n>p_{n-1}$). 
The asymptotic behaviors of the scaling function $f_c(z)$ are derived in Appendix B and read
\begin{eqnarray}
f_c(z) \to \begin{cases}
& 1+\frac{1}{{}_0F_1(1,-p_1^2/4)}\left(1-\frac{p_1^2}{4z}\right)e^{-p_1^2/(4z)}\, \quad {\rm as}\quad z\to 0 \\
& \\
& \sqrt{2}\, z^2\, e^{-z}\, \hskip 4.1cm {\rm as}\quad z\to \infty \, ,
\end{cases}
\label{fc_asympt}
\end{eqnarray}
where we recall that ${}_0F_1(1,z)$ is the confluent hypergeometric function and  
$p_1=-3.83171\dots $ is the negative root of $J_1(p)$ closest to the origin. 
Thus, for $L<< L_c(t)=a\, t$,
the PDF attains its stationary power law distribution $\sim L^{-2}$, which
gets cut-off exponentially fast for $L>> L_c(t)=a\, t$. 
The front $L_c(t)=a\, t$ separating
these two behaviors moves ballistically with speed $a$. 
This scaling function $f_c(z)$, obtained via numerical simulations, is compared to
our analytical prediction in \eqref{exact_fc} in  
Fig.~\ref{fig.maximum_critical}, finding excellent agreement. Interestingly, the scaling function
$f_c(z)$ in Fig.~\ref{fig.maximum_critical} is non-monotonic as a function of $z$. As $z$ increases,
the scaling function increases initially, displays a peak
before starting to decay monotonically as $z$ increases further.

\begin{figure}
\includegraphics[width=0.8\textwidth]{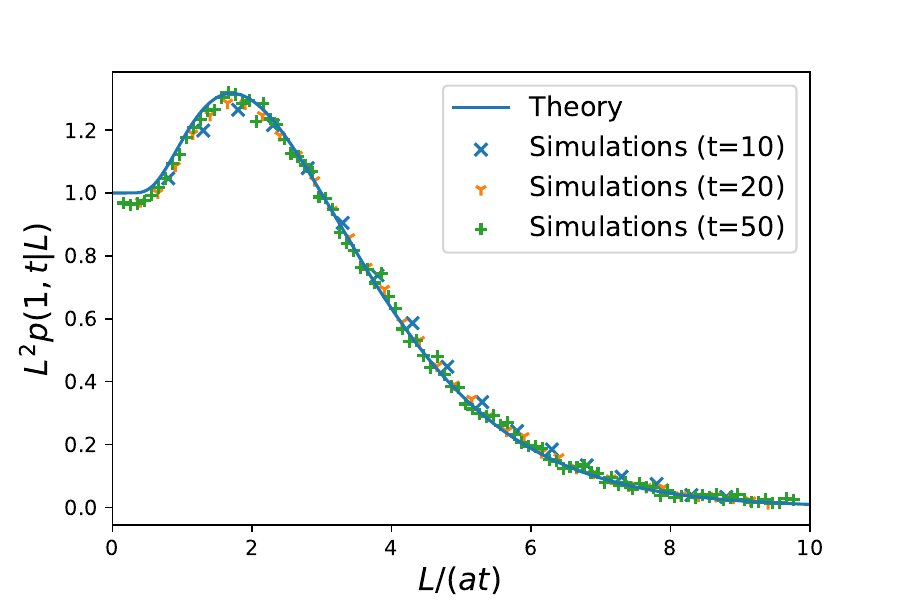}
\caption{Scaled probability distribution $L^2 p(1,t|L)$ of the (scaled) maximal population size 
$L/(at)$ up to time $t$ in the supercritical phase. The symbols correspond to numerical simulations 
with $a=b=1$, and different values of $t$. The continuous blue line corresponds to the exact 
scaling funcion $f_c(z)$ in Eq. (\ref{exact_fc}).
\label{fig.maximum_critical}}
\end{figure}

\begin{figure}
\includegraphics[width=0.8\textwidth]{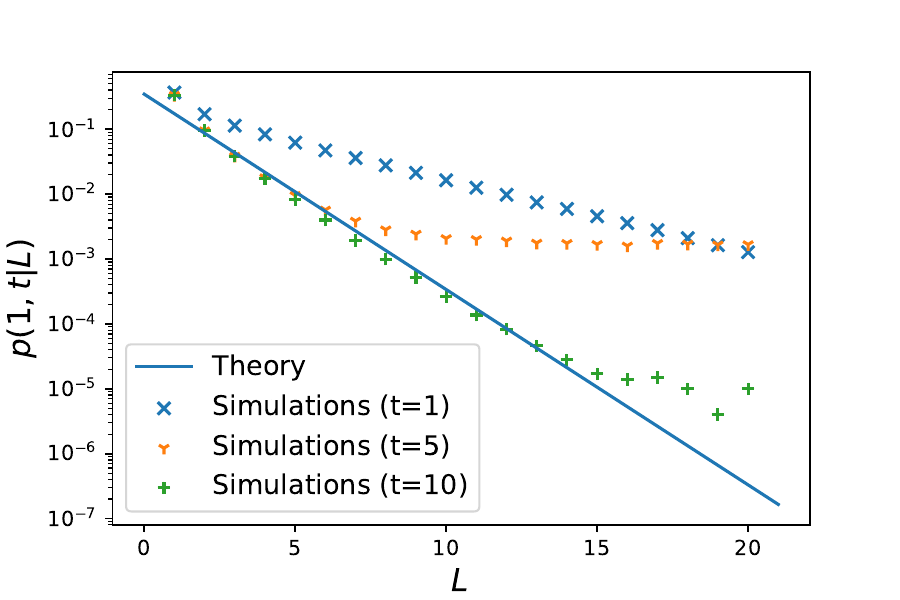}
\caption{Probability distribution $p(1,t|L)$ of the maximal 
population size $L$ up to time $t$ in the supercritical phase. 
The continuous blue line corresponds to the asymptotic 
result in Eq.~\eqref{super_pdf_final.1}, 
while the symbols correspond to numerical simulations with $a=1$, 
$b=2$, and different values of $t$. Note that the $\delta$-function 
in Eq.~\eqref{super_pdf_final.1} does not appear in the range of 
values shown in the figure.
\label{fig.maximum_supercritical}}
\end{figure}

\begin{figure}
\includegraphics[width=0.8\textwidth]{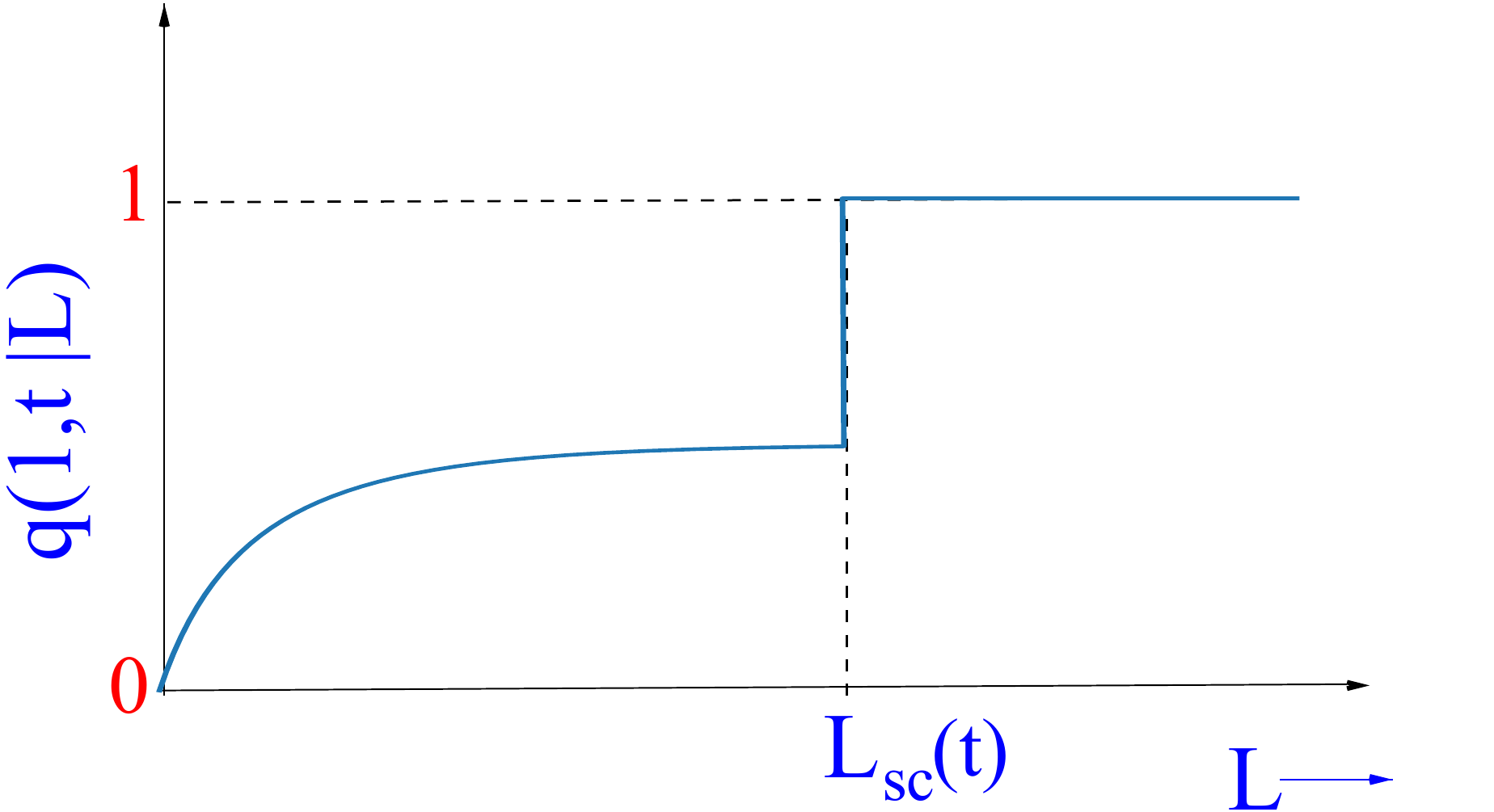}
\caption{A schematic plot of $q(1,t|L)$ vs $L$ in the supercritical 
phase $c<1$. The cumulative distribution grows with increasing $L$ saturating
to $c<1$ as $L\to  L_{\rm sc}(t) \approx e^{(b-a)t}$. When $L$ crosses 
$L_{\rm sc}(t)$, the cumulative distribution jumps to $1$.
}
\label{fig.q1t_super}
\end{figure}

\item {\bf Supercritical} phase $c=a/b<1$. In this case, as in the
subcritical phase, the integral $\tilde{J}_L(0)$ is finite and is given by
Eq. (\ref{sub_JL0.1}). Substituting this in Eq. (\ref{q1s.2})
and inverting the Laplace transform, we find that $q(1,t|L)$ again
approaches a time-independent form
\begin{equation}
q(1,t|L) \xrightarrow[t\to \infty]{} 1- \frac{1-c}{1-c^L} \, .
\label{super_q1t.1}
\end{equation}
However, unlike in the subcritical phase ($c>1$), 
in the supercritical phase where $c<1$, we find from Eq. (\ref{super_q1t.1}) that as $L\to \infty$
\begin{equation}
q(1,t\to \infty|L) \xrightarrow[L\to \infty]{} c<1 \, .
\label{super_norm.1}
\end{equation}
Thus, the corresponding time-independent PDF 
$p(1,\infty|L)=q(1,t\to \infty)-q(1,t|L-1)$ for large $L$ has an exponentially decaying tail
\begin{equation}
Q(L, t\to \infty)= p(1,\infty|L) \approx (-\ln c)\, (1-c)\, c^L \, ,
\label{super_pdf.1}
\end{equation}
and is not normalized to unity. In fact, the total probability weight $c<1$ carried
by this part of the PDF is precisely
the extinction probability $P(0,t\to \infty)$ in Eq. (\ref{p0t.1}).
Thus this part of the PDF of the maximum that becomes time-independent
at late times is associated to
samples or realizations of the process where the population becomes eventually extinct in the
supercritical phase. The rest of the PDF with mass $1-c$ is associated
with realizations in the supercritical phase where the population
explodes. We will call this part of the PDF as `condensate'.
 
To compute this condensate part, we need to analyse
$q(1,t|L)$ for finite but large $t$, which is done
in detail in Appendix C. We just summarize our main result here.
For finite but large $t$ and large $L$, keeping $t- \ln(L)/(b-a)$ fixed,
we simply obtain 
\begin{equation}
q(1,t|L) \approx 1- \frac{1-c}{1-c^L}\, 
\theta\left(t- \frac{\ln L}{b-a}\right)\, ,
\label{super_front.1}
\end{equation}
where $\theta(z)$ is the Heaviside theta function: $\theta(z)=1$ for $z>0$
and $\theta(z)=0$ for $z<0$.
From Eq. (\ref{super_front.1}), it follows that
\begin{eqnarray}
q(1,t|L) \to
\begin{cases}
& 1- \frac{1-c}{1-c^L} \quad {\rm as}\quad 
t\to \infty \quad {\rm for}\,\, {\rm fixed}\,\, L \\
& \\
& 1 \hskip 1.6cm {\rm as}\quad 
L\to \infty \quad {\rm for}\,\, {\rm fixed}\,\, t \, .
\end{cases}
\label{super_front.2}
\end{eqnarray}
Thus, there is a exponentially growing 
time-dependent scale $L_{\rm sc}(t)
\approx e^{(b-a) t}$ such that $q(1,t|L)$, as a function of $L$,
undergoes a discontinuous jump to $1$ from the value $1- (1-c)/(1-c^L)$
 as $L$ crosses $L_{\rm sc}(t)\approx e^{(b-a) t}$ (see Fig.
(\ref{fig.q1t_super}) for a schematic plot). Consequently,
taking a derivative of $q(1,t|L)$ with respect to $L$, we get
the PDF $p(1,t|L)$ of the maximum for large $L$ and $t$
\begin{eqnarray}
Q(L,t)= p(1,t|L)\approx \partial_L q(1,t|L)\approx (-\ln c)\, (1-c)\, c^L \theta\left(
e^{(b-a)t}-L\right) + (1-c)\, \delta\left(L- e^{(b-a)t}\right)\, .
\label{super_pdf_final.1}
\end{eqnarray}
The `delta' peak at the end of the regular (the second term) part
of the PDF corresponds precisely to the `condensate' that carries
an additional probability weight $(1-c)$ and corresponds to
realizations where the population explosion occurs. 
This exact expression in Eq.~\eqref{super_pdf_final.1} is shown in Fig.~\ref{fig.maximum_supercritical} 
and is in good agreement with numerical simulations for large $t$.

\end{itemize}

\section{Conclusion}
\label{conclu}

In this paper, we have studied the dynamics of the population size $N(t)$ of a branching process with death, where an 
individual splits into two daughter with rate $b$ and dies with rate $a$. Each individual may also diffuse as in branching 
Brownian motion with death, but the spatial diffusion does not affect the statistics of $N(t)$. While the instantaneous 
distribution $P(n,t)= Prob. [N(t)=n|N(0)=1]$ was well known, in this paper our main interest was on the history of the process 
$N(t)$ up to a fixed time $t$. In particular, we were interested in the extreme value statistics, i.e., the distribution of 
the maximal population size $M(t)= \max_{0\le \tau\le t}\,[N(\tau)]$ up to time $t$. We showed that this extreme value 
distribution can be computed exactly by mapping the nonlinear branching process into an auxiliary random walk problem on the 
positive semi-infinite axis, where the walker hops from site $n$ to one of its neighbours: to site $n+1$ with rate $b\, n$ and 
to site $n-1$ with rate $a\, n$. The process terminates when the walker hits the origin for the first time. This random walk 
representation with a state dependent hopping rate that increases linearly with distance from the origin is particular useful 
to obtain an exact solution of the problem for the distribution $Q(L,t)= {\rm Prob.}[M(t)=L]$ (starting at $N(0)=1$). The 
distribution $Q(L,t)$ exhibits markedly different bbehaviors in the subcritical ($b<a)$, critical ($b=a$) and the 
supercritical $(b>a)$ phases. In the subcritical phase, $Q(L,t)$ becomes stationary as $t\to \infty$ and this 
stationary distribution $Q(L, \infty)$ decays exponentially with increasing $L$. At the critical point $b=a$ also, the distribution 
again becomes independent of $t$ for large $t$, but $Q(L,\infty)\sim 1/L^2$ has a power law tail 
for large $L$. For finite but large $t$, the distribution at the critical point exhibits a scaling form $Q(L,t)\sim 
f_c(L/{at})/L^2$ with a nontrivial scaling function  $f_c(z)$ that we compute analytically. In the supercritical 
phase, the distribution $Q(L,t)$ has a `fluid' part that becomes independent of $t$ for large $t$ and a `condensate' part (a 
delta peak centered at $e^{(b-a)t}$) which gets disconnected from the fluid part and moves exponentially fast to $\infty$ as time 
increases.  Since the underlying process $N(t)$ is strongly correlated in time $t$, our results thus provide an exactly 
solvable case for the extreme value statistics for strongly correlated variables, a subject that has seen a lot of interests 
off late~\cite{EVS_review,EVS_book}. Besides, our results can be applied to characterize the
statistics of the maximally infected population up to a certain fixed time during the spread of an epidemic.

The population size $N(t)$ in our model at the critical point $b=a$ performs an unbiased random walk on the semi-infinite 
axis (with the origin as a sink) where the hopping rate out of a site at $n$ is proportional to $n$. Thus the random walker 
gets more and more `active' or `agitated' as it goes further and further away from the origin. This is very different from 
the `sluggish' random walk models studied recently in the literature in the context of slow subdifusive dynamics in 
inhomogeneous systems~\cite{Zodage2023,DV2025} and the stochastic porous medium equation~\cite{BFLR2025}, where the hopping 
rate out of a site at $n$ scales as $|n|^{-\alpha}$ with $\alpha>0$. In those models, the walker gets more and more 
sluggish as it deviates futher and further from the origin, leading to a very slow subdiffusive dynamics at late times. The 
behavior in our model at the critical point is thus quite different from this sluggish random walk model. A natural 
generalisation of this `agitated' random walk (ARW) model would be when the hopping rate out of a site at $n$ scales as 
$n^{\gamma}$ with $\gamma>0$. It would be nice to see if there is an underlying `branching' type process that can be 
represented by this generalised ARW model with index $\gamma>0$. In this paper, we have computed exactly the distribution 
of the maximum up to time $t$ in this ARW model with index $\gamma=1$. It would be interesting to compute the distribution 
of the maximum for the ARW model with arbitrary $\gamma>0$. Finally, an interesting extension of this problem would be to 
study this branching process in the presence of stochastic resetting~\cite{EM2011,EMS2020} whereby the population size 
$N(t)$ gets reset to its initial value $1$ with a constant rate $r$. For example, if a bacterial population is subject to 
antibiotics at random times, a large fraction dies thus effectively resetting the process. In this case, it would be 
interesting to study the fate of the population size of this branching-resetting process at long times.

\begin{acknowledgments}

We thank Francesco Mori for many useful discussions and for generously allowing us to use some of his simulations results.
SNM acknowledges support from ANR Grant No. ANR-23-CE30-0020-01 EDIPS.

\end{acknowledgments}

\appendix

\section{Derivation of the exact cumulative distribution of the maximum in the critical case $a=b$}
\label{A1}

In this appendix, we provide a derivation of the result in Eq. (\ref{q1s_crit.2}) in the critical
case $b=a$. Since the expression of $\tilde{J}_L(s)$ in Eq. (\ref{JLs.1}) is a bit singular when $b\to a$,
it is more convenient to take this limit directly in the first-order differential equation (\ref{wzs.1}) for
$W(z,s)$ and derive the result for the critical case. Setting $b=a$ in Eq. (\ref{wzs.1}), and solving
it we get
\begin{equation}
W(z,s)= A\, \frac{z}{(1-z)^2} \, e^{ \frac{s}{a\, (1-z)}}\, ,
\label{A1.1}
\end{equation}
where the unknown constant $A$ is fixed from the condition
\begin{equation}
\tilde{r}(L,s)=A\,  \oint \frac{dz}{2\pi\, i}\, \frac{1}{z^L\, (1-z)^2}\, e^{\frac{s}{a(1-z)}} = -\frac{1}{s}\, .
\label{A1.2}
\end{equation}
From the generating function $W(z,s)$, one can then obtain $\tilde{r}(n,s)$ and hence
$\tilde{q}(n,s)=1/s+\tilde{r}(n,s)$. Setting further $n=1$, we then obtain Eq. (\red{q1s.2})
for the critical case $b=a$ 
\begin{equation}
\tilde{q}(1,s)= \int_0^{\infty} q(1,t|L)\, e^{-s\, t}\, dt
= \frac{1}{s}\left[1-\frac{1}{ \tilde{J}_L(s)}\right]\, ,
\label{A1.3}
\end{equation}
with
\begin{equation}
\tilde{J}_L(s)= \oint \frac{dz}{2\pi\, i}\, \frac{1}{z^L\, (1-z)^2}\, e^{\frac{s}{a(1-z)}}\, .
\label{A1.4}
\end{equation}
To perform this integral, we make the change of variable
\begin{equation}
\frac{s\, z}{a (1-z)}= v \, ,
\label{A1.5}
\end{equation}
which leads to
\begin{equation}
\tilde{J}_L(s)= \frac{a}{s}\, \oint  \frac{dv}{2\, \pi\, i}\, \frac{1}{v^L}\, \left(\frac{s}{a}+v\right)^L \, e^v\, .
\label{A1.6}
\end{equation}
Using Cauchy's theorem, the integral in Eq. (\ref{A1.6}) is precisely the coefficient of $v^{L-1}$
in the power series expansion of $ (s/a+v)^L \, e^v$. Expanding this product in a power series in $v$ we get
\begin{equation}
\left(\frac{s}{a}+v\right)^L\, e^v= \sum_{n=0}^L {L\choose n}\, v^n\, 
\left(\frac{s}{a}\right)^{L-n}\, \sum_{m=0}^{\infty} \frac{v^m}{m!} \, ,
\label{A1.7}
\end{equation}
from wich one can read off the coefficient $C_{L-1}$ of $v^{L-1}$ as
\begin{equation}
C_{L-1}=  \sum_{k=1}^L {L\choose k}\, \frac{1}{(k-1)!}\left(\frac{s}{a}\right)^k= \frac{s}{a}\, L\, {}_1F_1\left(1-L,
2, -\frac{s}{a}\right)\, .
\label{A1.8}
\end{equation}
Substituting this result in Eq. (\ref{A1.6}) and then in (\ref{A1.3}) gives our desired 
result (\ref{q1s_crit.2}), i.e.,
\begin{equation}
\tilde{q}(1,s)\Big|_{b=a}=
\int_0^{\infty} q(1,t|L)\, e^{-s\, t}\, dt
= \frac{1}{s}\left[1-\frac{1}{L\,\, {}_1F_1\left(1-L,2,-\frac{s}{a}\right)}
\right]\,
\label{A1.9}
\end{equation}

\section{Derivation of the scaling function $f_c(z)$ in Eq. (\ref{crit_larget.1})}
\label{A2}

Let us first anticipate the scaling form in Eq. (\ref{crit_larget.1}) and then show that
it is true (aposteriori) by computing the scaling function $f_c(z)$. Assuming the
validity of Eq. (\ref{crit_larget.1}), we get the assocaited scaling form for the
cumulative distribution
\begin{equation}
q(1,t|L)= \int_0^L p(1,t|L') \, dL' =  1- \int_L^{\infty} p(1,t|L') dL' 
\approx  1- \int_L^{\infty} \frac{1}{L'^2}\,
f_c\left(\frac{L'}{a\, t}\right)\, dL' = 1- \frac{1}{a\, t}\, F_c\left(\frac{L}{a\, t}\right)\, 
\label{A2.2}
\end{equation} 
where
\begin{equation}
F_c(z)= \int_z^{\infty} \frac{f_c(x)}{x^2}\, dx \, \quad {\rm or\,\,\, equivalently}\quad 
f_c(z)= - z^2\, \frac{dF_c(z)}{dz}\, .
\label{A2.3}
\end{equation}
Hence if we can determine $F_c(z)$, we can extract the desired scaling function $f_c(z)$ from Eq. (\ref{A2.3})
Substituting the scaling form (\ref{A2.2}) for $q(1,t|L)$ in is exact Laplace transform in Eq. (\ref{q1s_crit.2}),
we get
\begin{equation}
\int_0^{\infty} \frac{dt}{a\, t}\, F_c\left(\frac{L}{a\, t}\right)\, e^{-s\,t}
\approx \frac{1}{s\,L}\, \frac{1}{ {}_1F_1\left(1-L, 2, -\frac{s}{a}\right)}\, ,
\label{A2.4}
\end{equation}
where we recall the definition of the hypergeometric function
\begin{equation}
{}_1F_1(\alpha,\beta,z)= 1+ \frac{\alpha}{\beta}\, x+ \frac{\alpha(\alpha+1)}{\beta(\beta+1)}\, \frac{x^2}{2!}
+\ldots\, .
\label{A2.5}
\end{equation}
Hence 
\begin{equation}
{}_1F_1\left(1-L,2, -\frac{s}{a}\right)= 1 + \frac{(L-1)}{2}\, \left(\frac{s}{a}\right)+ 
\frac{(L-1)(L-2)}{3!\, 2!}\, \left(\frac{s}{a}\right)^2+ \ldots
\label{A2.6}
\end{equation}
We now take the scaling limit $s\to 0$, $L\to \infty$, keeping
the product $s\,L$ fixed. Taking this limit term by term in the series in Eq. (\ref{A2.6}), we get
\begin{equation}
{}_1F_1\left(1-L,2, -\frac{s}{a}\right)\to \sum_{m=0}^{\infty} \frac{1}{m! (m+1)!}\,\left(\frac{sL}{a}\right)^{m}
= \sqrt{\frac{a}{sL}}\, I_1\left( 2\, \sqrt{\frac{sL}{s}}\right)\, ,
\label{A2.7}
\end{equation}
where $I_1(z)$ is the modified Bessel function. Finally, substituting Eq. (\ref{A2.7}) on the rhs
of Eq. (\ref{A2.4}), and setting $p= sL/a$, we get in the scaling limit (with $p$ fixed) the exact
relation satisfied by the scaling function 
\begin{equation}
\int_0^{\infty} \frac{dy}{y}\, F_c\left(\frac{1}{y}\right)\, e^{-p\, y}= \frac{1}{\sqrt{p}\, 
I_1\left(2\, \sqrt{p}\right)}\, .
\label{A2.8}
\end{equation}
Defining
\begin{equation}
\frac{1}{y} F_c\left(\frac{1}{y}\right)= G_c(y)\, , \quad {\rm or\,\,\, equivalently}\quad F_c(z)= \frac{1}{z}\, G_c\left(\frac{1}{z}\right)\,
\label{A2.9}
\end{equation}
one then has the exact Laplace transform for the function $G_c(z)$ from Eq. (\ref{A2.8})
\begin{equation}
\int_0^{\infty} G_c(y)\, e^{-p\, y}\, dy= \frac{1}{\sqrt{p}\,
I_1\left(2\, \sqrt{p}\right)}\, .
\label{A2.10}
\end{equation}

To proceed, we first formally invert Eq. (\ref{A2.10}) using
the Bromwich inversion formula
\begin{equation}
G_c(y)= \int_{\Gamma} \frac{dp}{2\,\pi\, i}\, \frac{e^{p\, y}}{\sqrt{p}\, I_1\left(2\sqrt{p}\right)}\, ,
\label{A2.12}
\end{equation}
where $\Gamma$ denotes the vertical Bromwich contour in the complex $p$-plane whose real coordinate
lies to the right of all singularities of the integrand. To evaluate this complex integral, we first identify the poles of the integrand, which has a pole at $p_0=0$ and infinitely many poles on the negative real line at $p_n$ with $n\geq 1$. The pole $p_n$ coincides with the $n$-th negative root of the Bessel function $J_1(p)$ (we will use the convention $p_n>p_{n+1}$). Evaluating the residues using Mathematica, we find that the function $G_c(y)$ can be written as
\begin{equation}
G_c(y)= 1+\sum_{n=1}^{\infty}\frac{1}{{}_0F_1(1,-p_n^2/4)}e^{-p_n^2y/4}\,,
\label{exact_Gc}
\end{equation}
where ${}_0F_1(1,z)$ is the confluent hypergeometric function. Moreover, the scaling function $f_c(z)$ can be written in terms of $G_c(z)$ as
\begin{equation}
f_c(z)= -z^2\, \frac{dF_c(z)}{dz}= G_c\left(\frac{1}{z}\right)+ \frac{1}{z}\, G_c'\left(\frac{1}{z}\right)\, .
\label{A2.11}
\end{equation}
Plugging the exact expression in Eq.~\eqref{exact_Gc} into Eq.~\eqref{A2.11}, we find
\begin{equation}
f_c(z)= 1+\sum_{n=1}^{\infty}\frac{1}{{}_0F_1(1,-p_n^2/4)}\left(1-\frac{p_n^2}{4z}\right)e^{-p_n^2/(4z)}\,.
\label{exact_fc_appendix}
\end{equation}

We can also extract the asymptotic behaviors of $f_c(z)$ for  $z\to 0$ and $z\to \infty$.
\begin{itemize}
\item {\bf The limit $z\to 0$} In this limit, the scaling function $f(z)$ in Eq.~\eqref{exact_fc_appendix} goes to the constant value $f(0)=1$. The leading order correction to this value for small but finite $z$ is given by the first term ($n=1$) in the sum in Eq.~\eqref{exact_fc_appendix}, yielding 
\begin{equation}
f(z)\approx 1+\frac{1}{{}_0F_1(1,-p_1^2/4)}\left(1-\frac{p_1^2}{4z}\right)e^{-p_1^2/(4z)}\,,
\label{exact_fc_appendix.1}
\end{equation}
where $p_1=3.83171\ldots$ and $p_1^2/4=3.67049\ldots$

\item {\bf The limit $z\to \infty$.} To evaluate the behavior of $f_c(z)$ for large $z$, it is useful to compute the asymptotic behavior of $G_c(y)$ in Eq.~\eqref{A2.12} for small $y$. In this limit, the dominant contribution to the integral comes from the large $p$ regime of the integrand. Using the leading asymptotic behavior
\begin{equation}
I_1(z) \to \frac{ e^{z}}{\sqrt{2\, \pi\, z}} \quad {\rm as}\quad z\to \infty\, ,
\label{A2.13}
\end{equation}
we get from Eq.~(\ref{A2.12})
\begin{equation}
G_c(y) \approx \int_{\Gamma} \frac{dp}{2\,\pi\, i}\, p^{-1/4}\, e^{p\, y- 2\, \sqrt{p}}\, .
\label{A2.14}
\end{equation}
We next rescale $p= s/y^2$ that gives
\begin{equation}
G_c(y) \approx \frac{2\sqrt{\pi}}{y^{3/2}}\, \int_{\Gamma} \frac{ds}{2\,\pi\, i}\ s^{-1/4}\,
e^{\frac{1}{y}\, (s-2\sqrt{s})}\, .
\label{A2.15}
\end{equation}
For small $y$, one can then evalutate this integral by the saddle point method that gives
\begin{equation}
G_c(y) \xrightarrow[y\to 0]{} \frac{\sqrt{2}}{y} \, e^{-1/y}\, .
\label{A2.16}
\end{equation}
Thus as $y\to 0$, the function $G_c(y)$ vanishes extremely rapidly with a leading essential singular
behavior $e^{-1/y}$. Plugging this behavior into the relation in Eq.~\eqref{A2.11}, we find
\begin{equation}
f_c(z)\approx \sqrt{2}\, z^2\, e^{-z}\,, \quad {\rm as}\quad z\to \infty\, .
\end{equation}

\end{itemize}

Hence, summarizing the asymptotic behaviors of $f_c(y)$
 \begin{eqnarray}
f_c(z) \to \begin{cases}
& 1+\frac{1}{{}_0F_1(1,-p_1^2/4)}\left(1-\frac{p_1^2}{4z}\right)e^{-p_1^2/(4z)}\, \quad {\rm as}\quad z\to 0 \\
& \\
& \sqrt{2}\, z^2\, e^{-z}\, \hskip 4.1cm {\rm as}\quad z\to \infty \, ,
\end{cases}
\label{A2.23}
\end{eqnarray}
where we recall that ${}_0F_1(1,z)$ is the confluent hypergeometric function and  $p_1=-3.83171\dots $ is the negative root of $J_1(p)=0$ closest to the origin.

\end{document}